# Interactive molecular dynamics in virtual reality from quantum chemistry to drug binding: An open-source multi-person framework

Michael O'Connor,[1,2,3] Simon J. Bennie,[1,3] Helen M. Deeks,[1,2,3] Alexander Jamieson-Binnie,[1,3] Alex J. Jones,[1,2,3] Robin J. Shannon,[3] Rebecca Walters,[1,2,3] Thomas J. Mitchell,[1,4] Adrian J. Mulholland,[3] and David R. Glowacki[1,2,3*]

[1]Intangible Realities Laboratory, University of Bristol, Cantock's Close, Bristol BS8 1TS, UK; [2]Dept. of Computer Science, University of Bristol, Merchant Venturer's Building, Bristol BS8 1UB, UK; [3]Centre for Computational Chemistry, School of Chemistry, University of Bristol, Cantock's Close, Bristol BS8 1TS, UK; [4]Creative Technologies Laboratory, University of the West of England, Bristol, BS16 1QY, UK

*glowacki@bristol.ac.uk

## Abstract

As molecular scientists have made progress in their ability to engineer nano-scale molecular structure, we are facing new challenges in *our ability to engineer molecular dynamics (MD) and flexibility*. Dynamics at the molecular scale differs from the familiar mechanics of everyday objects, because it involves a complicated, highly correlated, and three-dimensional many-body dynamical choreography which is often non-intuitive even for highly trained researchers. We recently described how interactive molecular dynamics in virtual reality (iMD-VR) can help to meet this challenge, enabling researchers to manipulate real-time MD simulations of flexible structures in 3D. In this article, we outline various efforts to extend immersive technologies to the molecular sciences, and we introduce 'Narupa', a flexible, open-source, multi-person iMD-VR software framework which enables groups of researchers to simultaneously cohabit real-time simulation environments to interactively visualize and manipulate the dynamics of molecular structures with atomic-level precision. We outline several application domains where iMD-VR is facilitating research, communication, and creative approaches within the molecular sciences, including training machines to learn reactive potential energy surfaces (PESs), biomolecular conformational sampling, protein-ligand binding, reaction discovery using 'on-the-fly' quantum chemistry, and transport dynamics in materials. We touch on iMD-VR's various cognitive and perceptual affordances, and how these provide research insight for molecular systems. By synergistically combining human spatial reasoning and design insight with computational automation, technologies like iMD-VR have the potential to improve our ability to understand, engineer, and communicate microscopic dynamical behavior, offering the potential to usher in a new paradigm for engineering molecules and nano-architectures.

# 1. Introduction

In 1977, artificial and augmented reality pioneer Myron Krueger began his paper *Responsive Environments* with the observation that "human-machine interaction is usually limited to a seated [person] poking at a machine with [their] fingers or perhaps waving [their] hands over a data tablet." [1] Krueger went on to speculate that real-time, multi-sensory interaction between humans and machines might enable exciting and efficient new approaches for exploring realities that are otherwise impossible to access. Nanoscale molecular objects offer fertile testbeds for exploring new technological frontiers in human-computer interaction (HCI), owing to the fact that molecules represent objects that are important to society and industry, but which we are unable to directly perceive, and which are characterized by considerable three-dimensional dynamic complexity. As Krueger observed, the sensory modes we use to obtain insight and navigate the complex and dynamic *terra incognita* of nanoscale structures are limited: our representational methods are confined mostly to 2d, and primarily designed for parsing by our visual cortex (plots, images, movies, articles, etc.). Recent research in psychology and neuroscience has shown that our attention is enhanced when we engage in multi-sensory processing,[2,3] simultaneously integrating complex data across our various sensory channels, spanning the visual, auditory, olfactory, and somatosensory cortexes. In some sense, we do not make full use of the array of sensory, perceptual, and information processing machinery which we have evolved as thinking and feeling beings to make sense of the natural world around us. This not only limits our ability to understand the 3D complexity of dynamical microscopic systems; in many cases it is also extremely inefficient. Beyond a relatively small size threshold of ~50 atoms, 2D representational tools quickly become unwieldy for handling 3D molecular systems. For example, researchers lose lots of time struggling with 2D molecular viewers to build complex 3d structures, attempting to represent 3D structural dynamics in a 2D presentation format, or fighting with scripting languages to undertake complex 3D molecular manipulations.

Over the past several years, our laboratory has been carrying out an interdisciplinary research program exploring interactive molecular dynamics (iMD) beyond standard 2d interfaces, designed to enable direct multisensory interaction with molecular simulations.[4-9] The recent emergence of robust and affordable virtual reality (VR) technologies has been a key enabler in these efforts, allowing us to develop a framework where scientists can manipulate rigorous real-time simulations of molecular systems, as shown in Figure 1 and **Video 1** (vimeo.com/244670465). Figure 1 shows two optically tracked participants (each wearing a VR head-mounted display (HMD) and holding in each hand wireless controllers which function as atomic 'tweezers') manipulating the real-time MD of a $C_{60}$ molecule using interactive molecular dynamics in virtual reality (iMD-VR). As shown in the video, the participants can easily 'lock onto' individual $C_{60}$ atoms and manipulate their real-time dynamics to pass the $C_{60}$ back and forth between one other. This is possible and immediately intuitive because interaction with the real-time $C_{60}$ simulation and its associated ball-and-stick visual representation is perfectly co-located – i.e., the interaction site in 3D physical space is exactly the interaction site in 3D simulation space. The client/server architecture illustrated in Figure 1 provides each VR client access to global position data of all other participants, so that any participant can see through their headset a co-located visual representation of all other participants. To date, our available resources and space constraints have allowed us to simultaneously co-locate six participants in the same room within the same simulation. The interaction shown in **Video 1**, where multiple participants in the same room are able to easily pass a simulated molecule between themselves (or e.g., collaboratively tie a knot in a protein) as if it were a tangible object, represents a class of simulated virtual experience which is simply not possible within the large-scale immersive stereoscopic CAVE environments that have become popular within academic and industrial research institutions.[10] In previous work,[8] we have shown that – with an good network connection – the Figure 1 MD server can be cloud-mounted, so that remotely located workers can occupy the same virtual space.

Whilst adoption of new forms of immersive human computer interaction like iMD-VR is not yet widespread within the molecular sciences, several case studies across a range of fields have demonstrated *the quantitative benefits* that arise from utilizing immersive forms of human computer interaction beyond standard 2d GUIs and text-based displays. For example, in the medical field, detailed interactive surgical simulations in VR have an established track record for more than a decade. A number of studies have quantitatively shown that VR-trained surgeons complete surgical procedures faster, with significantly lower error rates (for example, a 2002 paper reported 7x fewer errors).[11] Similarly, digital animation firms like Dreamworks have reported time and cost reductions on the order of 3x following adoption of immersive technologies which allow their digital animators to reach into scenes and carry out direct manipulations (e.g., to animated characters) in 3d.[12] Recent controlled studies carried out in our lab have quantitatively demonstrated iMD-VR enables researchers to complete molecular modelling tasks more quickly (2x – 10x) than they can using conventional interfaces like a mouse or a touchscreen. The observed accelerations become increasingly significant for molecular pathways and structural transitions whose conformational choreographies are intrinsically 3-dimensional.[8] For reactive systems where multiple competing reaction channels are available, we have also recently demonstrated that iMD-VR combined with 'on-the-fly' *ab initio* quantum chemistry offers an extremely efficient strategy for sampling reactive geometries along the minimum energy path (MEP) which can be used to train neural networks.[13]

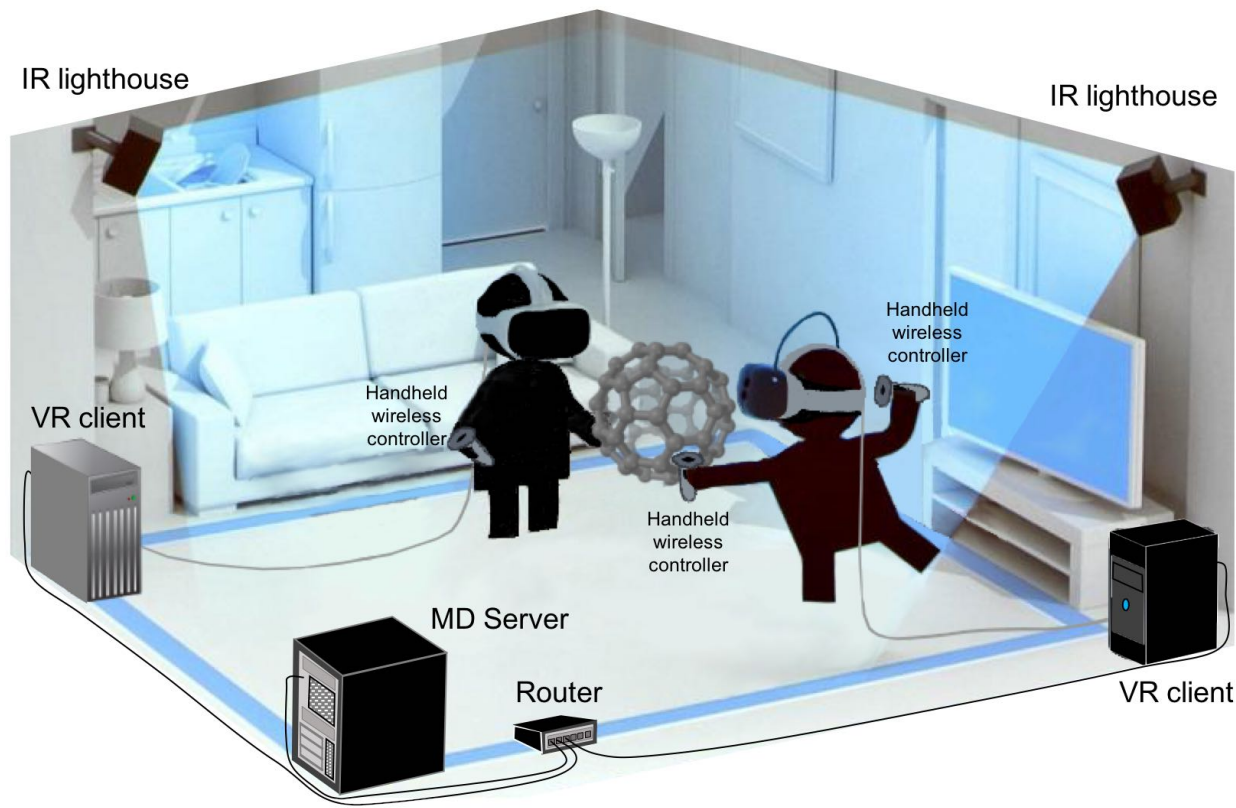

Figure 1: schematic of the setup for Narupa, our open-source multi-person iMD-VR framework, showing two participants using handheld wireless controllers to manipulate a real-time MD simulation of a $C_{60}$ molecule. The position of each user's head mounted display (HMD) and wireless controllers is determined using a real-time optical tracking system composed of synchronized IR light sources ('lighthouses'). Each user's HMD is rendered locally on a VR client computer fitted with a suitable GPU. MD calculations and maintenance of global user position data take place on a separate MD server, shown here as a local workstation (which alternatively can be cloud-mounted). As long as the network connecting client and server enables sufficiently fast data transfer, system latency is imperceptible. The figure shows a local network setup optimized to minimize latency, with the MD server and VR clients communicating via LAN cables connected to a router.

Beyond these quantitative benefits, *we have also observed a number of qualitative benefits* from iMD-VR. Whilst these benefits are less easy to capture in a graph, a table, or a number, they are nonetheless significant, because they enable improved understanding and insight into complex molecular systems, furnishing an improved sense for how molecular objects move and respond to perturbation, facilitating efficient clear communication, and encouraging researchers to think creatively about their systems. Many of the difficulties that arise in understanding and communicating aspects of molecular science result from the fact that our 'ways of knowing' the nanoscale molecular world are *indirect* – a result of the fact that molecular lengthscales are very small and molecular timescales are very fast. Unable to directly perceive these lengthscales and timescales using our human sensory apparatuses, we rely instead on instrumental data-feeds[14] and abstract models. Our ability to 'know' the molecular world relies primarily on our ability to parse instrumental data feeds and our ability to undertake cognitive abstraction and develop models, in order to make connections between abstract domains and experimental domains – e.g., using abstract models to generate an experimental hypotheses, and using experimental data to refine abstract models. Successful organic chemists, for example, are able to undertake complex cognitive mappings to translate between 3d molecular structures and their 2d molecular notation.

In his visionary essay *The Ultimate Display*, Ivan Sutherland highlighted our lack of intuition for scientific domains where we cannot directly perceive our objects of study and are therefore always 'one step removed': *We live in a physical world whose properties we have come to know well through long familiarity. We sense an involvement with this physical world which gives us the ability to predict its properties well. For example, we can predict where objects will fall, how well-known shapes look from other angles, and how much force is required to push objects against friction... We lack corresponding familiarity with the forces on charged particles, forces in non-uniform fields… and high-inertia, low friction motion…*[15] Over the last few decades, advances in experimental techniques like super-resolution fluorescence microscopy and cryo-electron microscopy have helped refine our ability to map molecular structure and dynamics. It has been proposed that advances in nano-engineering may one day allow us to design and construct nanoscale structures and machines with the sort of precision that is possible in the design and engineering of macroscopic objects. For example, in his oft-quoted 'plenty of room at the bottom' lecture,[16] Richard Feynman speculated that we would one day be able to carry out routine atomic level manipulation at the scale of individual atoms[17, 18] – a kind of atomically resolved surgery which remains a holy grail for scientists working at the nanoscale.

As we make progress in our ability to engineer and design molecular structure and function, a new fundamental challenge is emerging: *namely, our ability to understand and engineer molecular motion, dynamics, and flexibility*. Here we encounter exactly the sorts of 'non-intuitive' physics highlighted by Sutherland: forces acting on charged particles in non-uniform fields are the norm, and high-inertia-low-friction regimes are not uncommon. Moreover, because molecular systems typically have thousands of degrees of freedom, their motion is characterized by a complex, highly correlated, many-body dynamical choreography which is unintuitive because it has few analogues in our day-to-day experience. As we aim to not only engineer structure, but also dynamics and flexibility, we require tools which enable us to obtain a designer's sense of the properties of the various materials which constitute our building blocks. In the same way that VR enables better surgical peformance, we can imagine a scenario where biomimetic molecular designers use iMD-VR to gain a sense for how biological molecules move and how they 'feel', in order to make more informed creative design hypotheses.

In classifying different forms of human-computer-interaction, experts often refer to the 'affordances' of a particular environment or technology – i.e., the features of a particular environment or technology that elicit a particular kind of behavior or interaction.[19-22] For example, a computer screen-mouse-keyboard combination clearly has a distinct set of design affordances compared to a virtual reality interface. Both technologies enable the rendering of computer generated images; however, a virtual reality interface allows one to walk around in space to inspect the image from various angles and quickly intuit depth, while a screen requires that the user observe the image from a particular perspective and carry out a sequence of 2d mouse manipulations to understand depth. As another example, a keyboard primarily emphasizes the transmission of text-based information via button presses and a mouse affords one-handed manipulations in two dimensions in order to navigate the screen. Neither a set of wireless tracked VR controllers (like those shown in Figure 1) nor a pair of VR gloves is well suited to rapid text input like that which is afforded by a keyboard, but they afford precise and intuitive two-handed spatial manipulation. From a research perspective, a key question for the molecular sciences involves understanding those particular areas where the affordances of new VR environments (compared to 2d screen-mouse-keyboard environments) enable deeper insight, a better feel for nanoscale design and engineering, more effective scientific communication and collaboration, and accelerated research progress in understanding important molecular systems and concepts. [23] In a recent paper, Goddard *et al.* have outlined a number of the software tools which have emerged for use in head-mounted virtual reality environments, [24] many of which have their conceptual origins in software frameworks originally designed for use in stereoscopic, multi-projector CAVE-like environments. [10, 25-30] In the last few years, software frameworks which have emerged for head mounted VR displays can be broadly schematized according to the extent of active participation which they enable. These include applications:

- Enabling a participant to inspect either a static molecular structure or a pre-recorded molecular trajectory in three dimensions. In such applications, the role of the participant is primarily observational; the head mounted display essentially operates as a mechanism for enabling a 360-degree video where the participant can look around;[31-34]
- Where a participant has a more active role, and can navigate a simulated space to inspect a structural rendering from various angles and quickly intuit depth. In many cases, participants are able to manipulate aspects of the structural model or

trajectory – e.g., changing its representation and rendering options, pausing and resuming the trajectory, showing or hiding certain parts of the structure, rotating/translating the model, and perhaps querying structural aspects of the model (e.g., bond distances, angles, residue names, etc.); [24, 35-39]

- Enabling a participant to carry out modifications on a molecular structure, e.g., to build or modify molecules by connecting together atoms or amino acids, replacing one functional group with another functional group, etc.;[35]

To date, our own research has specifically explored interactive molecular dynamics in virtual reality (iMD-VR) – i.e., applications that emphasize simulation at interactive latencies, in which the affordances of two-handed interaction within the three-dimensional VR space enable a participant to 'reach out and manipulate' rigorous MD simulations, and carry out detailed three-dimensional structural manipulations in real-time, as shown in Figure 1.[8] Our experiments suggest that the utility of iMD-VR as a research tool arises from its ability to transform abstract molecular models into tangible dynamic realities. In section 2 of this article, we discuss VR's recent resurgence, and outline a particularly useful way to schematize the various types of VR which are emerging in the commodity market. We follow this with a brief history of interactive molecular simulation, and then discuss participants' reports that they can 'feel' the dynamics of simulated molecular objects within the iMD-VR environment. Section 3 of this article outlines a project we have provisionally named 'Narupa', an open-source (GPL v3.0) iMD-VR software framework which we made publicly available to coincide with this article (source at gitlab.com/intangiblerealities and a stable executable at irl.itch.io/narupaxr). The name 'Narupa' combines the prefix 'nano' and suffix 'arūpa' (a Sanskrit word describing non-physical and non-material objects), which represents our attempt to capture what it is like to interact with simulated nanoscale objects in VR. Section 4 details a variety of research applications which we have carried out using Narupa, and section 5 outlines some new interaction strategies which are exploring in order to increase the utility of iMD-VR for researchers in the molecular sciences. In section 6, we conclude this article, and discuss research directions moving forward.

In designing Narupa, we have aimed at a 'low threshold-high ceiling' design.[40, 41] This established design paradigm emphasizes tools that are easy for novices to get started using (i.e., a low threshold), but which are sufficiently powerful to enable experts to make progess on sophisticated and complex projects (a high-ceiling). In the last couple years, we have made countless demonstrations of Narupa to international research colleagues, and we have observed that the 'low-threshold' aspect of our design approach leads some researchers to dismiss the software as 'a gimmick' or 'a toy'. At its core, however, Narupa is a rigorous scientific simulation environment, with three key emphases: (1) the integration of real-time simulation methodologies into our interaction framework, enabling participants to manipulate and 'feel' the dynamical responses of molecular systems; (2) the ability to make the VR experience one in which facilitates communication, by enabling multiple participants to cohabit the same virtual world together, either together in the same room, or distributed remotely; and (3) active engagement with designers, artists, and human-computer interaction (HCI) experts, in order to create a framework which not only has scientific utility, but which is aesthetically compelling.[5, 7, 42] This latter point is particularly important given the level of immersion which can be achieved in VR environments. An unattractive aesthetic makes participants unlikely to utilize immersive tools.

Faced with traditional scientific publication formats, one of the most well-known difficulties for workers in VR concerns exactly how to write about it.[43] This a particularly important point for the purposes of this article, given that 'reaching out to touch molecules' falls into a class of perceptual experience which does not have a very good analogue in our day-to-day phenomenological experience. In an attempt to deal with this difficulty, this article makes reference to a number of videos (listed in Table 1), each with a hyperlinked URL, which we encourage the reader to watch alongside the corresponding text, because we have found that they go a long way toward overcoming the difficulties in writing about aspects of multi-person iMD-VR that are difficult to communicate with text alone.

| Video Index | URL | Description | Force Engine |
|---|---|---|---|
| Video 1 | vimeo.com/244670465 | Multi-person demo showing:[8]<br>(a) $C_{60}$ being passed back and forth;<br>(a) $CH_4$ transit through a nanotube;<br>(b) helicene changing from right to left-handed twist;<br>(c) 17-ALA peptide being tied in a knot | (a) MM3[44]<br>(b) MM3[44]<br>(c) MM3[44]<br>(d) OpenMM[45] |
| Video 2 | vimeo.com/305459472 | Illustrating the iMD-VR selection interface with Cyclophilin A | OpenMM[45] |
| Video 3 | vimeo.com/315239519 | Narupa secondary structure visualization demo of neuraminidase (PDB 3TI6) | OpenMM[45] |
| Video 4 | vimeo.com/315218999 | Reactive & non-reactive $OH + CH_4$ scattering | DFTB+[46] |
| Video 5 | vimeo.com/311438872 | Exploring reactive PESs for CN + isobutane for NN fitting using interactive *ab initio* quantum chemistry[13] | SCINE[47] |
| Video 6 | vimeo.com/312963823 | On-the-fly reaction discovery for OH + propyne using interactive *ab initio* quantum chemistry | SCINE[47] |
| Video 7 | vimeo.com/306778545 | Reversible Loop Dynamics in Cyclophilin A | OpenMM[45] |
| Video 8 | vimeo.com/274862765 | Using the Narupa-OpenMM plugin to dock benzamindine with trypsin | OpenMM[45] |
| Video 9 | vimeo.com/296300796 | Using the Narupa-OpenMM plugin to dock oseltamivir with neuraminidase | OpenMM[45] |
| Video 10 | vimeo.com/312957045 | Guiding 2-methyl-hexane through a ZSM-5 zeolite using the Narupa-PLUMED interface | PLUMED/DL_POLY[48, 49] |
| Video 11 | vimeo.com/312994336 | Real-time sonification of a biomolecule's potential energy illustrated by tying a knot in 17-ALA peptide | OpenMM[45] |
| Video 12 | vimeo.com/305823646 | Use of our custom Extextile VR gloves to tie a knot in 17-ALA peptide[50] | OpenMM[45] |

Table 1: videos discussed in this article, along with their respective URLs, a brief description, and the force engine utilized in the video

## 2. iMD and VR: context & history

*2.1 A hierarchy for classifying VR technologies*

A detailed review of the history of VR is available elsewhere. [43, 51] Whilst VR technologies have been available for much longer than the latest hype cycle, the distinguishing feature of the current resurgence is the fact that technology which was previously only available in specialist research labs or medical school facilities is now available at considerably lower prices. Driven mostly by the consumer gaming and entertainment market, recent advances in VR hardware provide commodity-priced solutions to the longstanding problem of co-located interaction in three dimensions. *HCI technologies are co-located when there is a perfect alignment between the interaction sites in physical space and the interaction sites in virtual space.* [52] Touchscreens, for example, solve the problem of 2D co-location because the interaction site in physical space is identical to the interaction site in virtual space. This is a significant reason why children at a very young age find it straightforward to navigate a touchscreen. Combining infrared optical tracking, inertial movement units (IMUs), and application specific integrated circuits (ASICS), commodity VR technology such as the HTC Vive offers fully co-located interaction in three dimensions, tracking a participant's real-time 3D position with errors less than a centimeter, and allowing participants to reach out and touch simulated objects in the virtual world, as shown in Figure 1 & **Video 1**.

A wide array of relatively distinct technologies are currently available which are often referred to as 'virtual reality', each of which offers different affordances. However, it is important to address a widespread misconception: *strapping a screen to one's head implies nothing about the level of immersion the participant experiences*. VR pioneers like Jaron Lanier have emphasized this point, highlighting the fact that a number of frameworks which are often referred to as 'virtual reality' enable participants to do little more than 'just looking around in a spherical video'. [51] Lanier, along with other HCI researchers, has made a point to distinguish those technologies *which do afford* reaching out to touch the virtual world: *If you can't reach out and touch the virtual world and do something to it, you are a second class citizen within it... a subordinate ghost that cannot even haunt.* [51] From this point on in this article, we use the term 'virtual reality' specifically in reference to technologies like the HTC Vive and the Oculus Rift, whose design affordances enable one to 'reach out and touch' simulated realities. In an excellent recent review of virtual reality principles and applications,[43] Mel Slater highlights a useful way to schematize different VR technologies *according to the level of immersion which they offer*. VR technologies are ultimately sophisticated simulators, and therefore any VR technology's level of immersion can be defined relative to another VR technology by making a determination as to whether its affordances enable it to simulate in principle (or not) the experiences available with another technology. So we can say that a specific VR technology A is 'more immersive' than another VR technology B so long as A could be designed (in principle) to simulate the experience of using B. Our efforts to date have focused primarily on the HTC Vive, because it represents one of the most immersive commodity frameworks according to this definition – i.e., its versatility enables it to simulate the vast majority of other VR technologies (e.g., a CAVE, [10] a Samsung Gear headset, a Playstation headset, etc.), but not *vice versa*.

Beyond virtual reality, other forms of technology are emerging which enable simulated immersive experiences, including augmented reality (AR), and mixed reality (MR). While a detailed discussion of these various emerging technologies is beyond the scope of this paper (and complicated owing to the fact that the technology is evolving rapidly), we note that the various forms of embodied digital interaction (whether they are forms of virtual, augmented, or mixed reality) are sometimes referred to on aggregate as 'XR', or 'extended' reality. Having experimented with a wide range of available technologies, we have found the aforementioned HTC Vive to be generally robust for the purposes of molecular simulation and visualization. Moreover, the HTC Vive also allows us to design experiences which enable groups of people within the same space (as shown in Figure 1) to simultaneously co-habit the same simulated virtual world. However, we note that the technology is steadily advancing, and many of the ideas in this paper are not limited to VR. They could easily be extended to any of a range of XR technologies, so long as their affordances enable one to 'reach out and touch' simulated realities, and then carry out spatial manipulations with a sufficient degree of precision so as to enable workers to carry out detailed atomic adjustments and rearrangements. Throughout this article, we describe people who use VR as 'participants' rather than 'users', recognizing that VR is different from other forms of human–computer interface because the human can actively *participate in the virtual world*. [43]

*2.2 Interactive Molecular Simulation*

Historical efforts to use computing to designing new ways to interact with molecular simulations have been strongly influenced by the kinds of tangible (e.g., plastic, metal, wood, etc.) molecular models that have been historically important in chemistry and biochemistry – e.g., tangible three-dimensional (3D) molecular models like Dorothy Hodgkin's crystallographic model of penicillin's structure, [53] Pauling's models to identify the structure of alpha-helices, [54] Watson and Crick's famous DNA model, and the [55] large room-sized models, made from e.g., wire, plastic, brass, balsawood, and plasticene which were used to refine and represent protein crystal structures by pioneers such as Kendrew and Perutz.[56, 57] Physical models like these provide structural insight, but cannot represent the often non-intuitive mechanics that determine how molecules move and flex. The first researchers to pursue the idea that computers could be used to construct tangible molecular models whose motion was based on rigorous physical laws included Fred Brooks[58] and Kent Wilson[59], pioneers whose interests spanned both scientific simulation and human-computer-interaction (HCI). Brooks and Wilson were amongst the first to imagine how such technology would offer better insight, and also have the potential to accelerate research workflows. Following on from the ideas outlined by Sutherland, they speculated that interactive molecular simulation (iMS) frameworks would lead to models which would be as intuitive to manipulate as the old tangible models, but which followed rigorous physical laws.

Brooks for example designed an immersive six-degree-of-freedom (DOF) force-feedback haptic system which participants could manipulate to carry out molecular tasks, [58, 60] mounted at the UNC Dept of Computer Science.[58, 60-62] The original system was built from an enormous robotic arm called the Argonne Remote Manipulator (ARM).[62] Building on evidence that force feedback tools allowed participants to efficiently carry out remote manipulation tasks relevant to space research, underwater operations, and nuclear/radiation laboratories, [63] Brooks sought to investigate whether the same was true for manipulation of molecular models. He designed a study in which participants were instructed to carry out a simple force minimization task emulating a ligand-receptor molecular docking-type problem – namely a rigid diatomic molecule in which each atom is acted upon by three unique harmonic forces, and initialized in a non-optimal configuration. His results suggested that participants were able to minimize the interaction potential energy faster relying upon "blind" force-feedback compared to visual feedback. Inspired by this work, Klaus Schulten and co-workers subsequently miniaturized Brooks' setup: by manipulating a desktop-mounted haptic pointer, participants could steer the real-time dynamics of molecules rendered on a stereographic screen. [64] This remains the approach utilized in most published iMS approaches[58] – i.e., the participant manipulates a small pen-shaped mouse that can move in three translational dimensions (x,y,z), and three rotational dimensions ($r_x$, $r_y$, $r_z$). This pen-shaped mouse is attached to a robotic arm which can be programmed to 'resist', a phenomenon which workers in HCI often refer to as 'force-feedback'. This approach has been extended by others, including Marc Baaden, Markus Reiher, Todd Martinez, and co-workers to interactively manipulate molecular mechanics [65] and quantum chemistry simulations.[66, 67]

*2.3 'Feeling' molecules in virtual reality*

The use of VR in surgical contexts – where it is intended to simulate a surgeon's experience of manipulating and cutting human tissues – is rather distinct from the use of VR to manipulate molecular structure and dynamics. Perhaps the most important difference pertains to the design reference. Surgical simulator applications have a well-defined and measurable design reference, with a well-defined design question: how does the simulation 'feel' compared to an experience involving human tissue? Molecular applications, on the other hand, have no similarly well-defined design reference – i.e., there is neither a clear answer to the question "What does a molecular system 'feel' like?" nor to the question "what *should* a molecular system 'feel' like?". The lack of reference is a central part of what makes developing a real-time molecular simulation and manipulation framework such a fascinating and creative challenge, which must necessarily consider aesthetics, design, and participant psychology in order to be effective.

Brooks did much to develop practical iMS strategies, and nearly everybody who has persisted in exploring iMS over the years has adopted his 6 DOF haptic approach. The miniaturization of such haptic devices has also made them practical for use within surgical simulators, where they can operate (for example) as a surgical knife, or be programmed to accurately simulate the resistance of tissues. As a result of the work by both Brooks and Wilson, many workers in iMS have concluded that simulating the 'feeling' of a molecular structure requires the use of force feedback haptics connected to robotic arms. One problem with these sorts of haptic devices is that they face a well-known limitation in their ability to achieve 3D 'co-location'. For interactive molecular simulations, 3D co-location is an important design consideration, owing to the fact that molecules are 3D objects which move in 3D. In principle, co-located solutions involving haptics are possible – e.g., by co-locating the haptic device within the VR environment. However, such strategies require compatibility between multiple layers of non-commodity technologies, and it remains unclear whether their technological cost and sophistication outweighs their benefits. Moreover, haptic technologies face fundamental limitations, owing to the fact that there are excellent solutions available for specific types of interaction (e.g., pushing a needle through tissue in a surgical simulation, or using an exoskeleton to apply force feedback to an arm); however, there are no *generalized* solutions in the form of a single device which enables participants in a VR environment to feel *anything* (e.g., in the same way that visual or auditory display can be programmed to display anything). For this reason, some have argued that a generalized haptic solution is likely only possible in the form of a direct brain interface, in which case haptics will becomes a branch of applied neuroscience.[68]

Haptic technologies like a robotic arm which a participant can pull, and which then pulls back, offer one particular form of 'felt' sensation; however, our own research experience to date strongly suggests that felt sensation can be accomplished without the use of haptic pointers. Indeed, our experience of taking thousands of people into iMD-VR over the past few years, and enabling them to manipulate a range of different molecular structures, has shown that people do indeed 'feel' molecular responses as they manipulate them. One particularly notable example of this occurred during an experiment which we will henceforth refer to as the 'Burke Perception Experiment' (BPE), carried out during a visit by Professor Kieron Burke to our lab in Bristol – and one particular comment which he made during his 30-minute iMD-VR experience. After Kieron successfully threaded methane through a nanotube, we instructed him to perturb a simulation of a small peptide (17-Alanine) from its native structure and then tie it into a knot. While manipulating the peptide, Kieron remarked "this feels so much different than the nanotube".

Kieron's comment during the BPE is notable because it is extremely common. Multiple people, from a wide range of backgrounds, *consistently* comment on the fact that different molecules simulations 'feel' differently. However, it is not obvious why these sorts of comments consistently emerge: watching people from outside of VR, as shown in **Video 1**, they appear to be grasping at air. They are not touching anything physical. In a first attempt to unravel the mechanisms which might be at play here, we have been developing a concept of 'layered perceptions'.[9, 69] At the moment, we believe that one's ability to 'feel' a molecular object in VR arises from a layering of visual perception on top of proprioception (the non-visual sense through which we perceive the position and movement of our body). So when Kieron Burke reaches out to 'touch' a nanotube in VR, he locks his force tweezers onto an atom (or selection atoms) in a nanotube, whose underlying physics are dominated by covalent interactions (simulated in real-time). The form of these forces requires Kieron to move in a particular way in order to make the system respond as he wishes. The protein, on the other hand, has dynamics which are largely governed by much weaker non-bonded interactions. And therefore, Kieron must move in a slightly different way in order to tie the protein into a knot. Our working hypothesis is that Kieron's proprioceptive sensations are working alongside his visual sense to project a sense of 'feeling' onto objects which are otherwise only virtual – i.e., his brain is integrating visual and proprioceptive details to 'fill in' the details of what such an object *would* feel like. This hypothesis is grounded in part from published work demonstrating that virtual reality can be used to heighten proprioceptive recovery in stroke patients,[70] along with research showing that well-constructed VR experiences operate so as to encourage the brain to 'fill in' the perceptual details of a given scenario.[43]

We are currently working to test these hypotheses in further detail. Whatever the precise mechanism, *it appears that people can 'feel' simulated objects which do not have a material essence*. That sense of 'feeling' provides them an embodied awareness of how nanoscale systems (simulated using classical dynamics on an approximate PESs) behave, and how they respond to perturbation. This is an important insight because it means that it is possible to 'feel' a molecule without expensive haptic technologies, which are non-commodity pieces of equipment and therefore tend to be rather expensive and cumbersome. Moreover, by heightening our proprioceptive sensitivities, it may be possible to enhance our ability to 'feel' simulated realities. Because there is no design reference for *what a molecule should feel like*, using subtle mechanisms like proprioception represent an approach which is equally reasonable compared to haptic approaches. The extent to which we can effectively design for the proprioceptive sense of feeling remains to be seen. Further HCI tests will provide insight into each approach's respective strengths and weaknesses.

## 3. An open-source multi-person iMD-VR environment

### 3.1 Narupa

Figure 1 and **Video 1** illustrates the Narupa framework we have developed to interface the HTC Vive with rigorous real-time molecular simulation algorithms, which we have released as an open-source project. Narupa overcomes several limitations of the proof-of-principle iMD-VR prototype framework we previously described in an article by O'Connor et al[8]: (a) it enables easy access to the multi-person functionality illustrated in Figure 1, so that multiple participants can inhabit the same iMD-VR environment; (b) rather than the simulations being predefined in advance, it enables participants to set up and customize their own simulations using a flexible force API (discussed further below); and (c) it can be set up to run on local networks (i.e., does not require access to cloud computing over fast networks). While a real-time MD simulation of $C_{60}$ using molecular mechanics force fields is relatively cheap, the client-server architecture shown in Figure 1 enables access to a more powerful computational back-end as needed, in order to simulate systems of increased complexity (discussed further in what follows). The URL irl.itch.io/narupaxr, where we have made the Narupa executable available, contains a link to documentation included as part of the open-source software repository, listing the hardware required to set up a multi-person VR environment which can accommodate $n$ participants (where $n \leq 8$), along with costs and instructions on how one goes about setting up a lab of their own.

### 3.2 Force biasing

The VR-enabled interactive MD shown in **Video 1** effectively amounts to a real-time classical dynamics simulation which responds to real-time biasing forces, building on our previous work using optical tracking technologies to interactively steer real-time molecular simulations. [4] In classical mechanics, the time-dependent dynamics of molecular systems are solved by numerically integrating Newton's equations of motion. The vector of forces acting on a set of atoms **F**($t$) can be written in terms of the system's potential energy $V$, i.e.:

$$\mathbf{F}(t) = -\frac{dV}{d\mathbf{q}} \qquad \text{Eq (1)}$$

where **q** is a vector containing the position of each atom in the ensemble. Our system effectively allows participants to interactively chaperone a real time MD simulation by splitting $V$ into two different components

$$V = V_{int} + V_{ext} \qquad \text{Eq (2)}$$

where $V_{int}$ corresponds to the system's internal potential energy, and $V_{ext}$ corresponds to the additional potential energy added when a participant exerts a force on a specific atom (or group of atoms) when they grab it using the handheld wireless controller shown in Figure 1. Substituting Eq (2) into Eq (3) then gives

$$\mathbf{F}(t) = -\frac{dV_{int}}{d\mathbf{q}} - \frac{dV_{ext}}{d\mathbf{q}} \qquad \text{Eq (3)}$$

The external forces can be implemented in a number of ways, including by projecting a spherical Gaussian field into the system at the point specified by the participant, and applying the field to 'lock onto' the nearest atom $j$ as follows:

$$\frac{dV_{ext}}{d\mathbf{q}} = \frac{m_j c}{\sigma^2}(\mathbf{q}_j - \mathbf{g_i})e^{\frac{-\|\mathbf{q}_j - \mathbf{g_i}\|^2}{2\sigma^2}} \qquad \text{Eq (4)}$$

where $m_j$ is the atomic mass of the nearest atom, $c$ is a scale factor that tunes the strength of the interaction, $\mathbf{q}_j$ is the position of atom $j$, $\mathbf{g}_i$ is the position of the interaction site, and $\sigma$ controls the width of the interactive fields. $c$ is variable parameter that the participant can set, so as to achieve responsive interaction while preserving dynamical stability, and $\sigma$ is typically set to the default value of 1nm. While an interaction is active, it is always applied to the same atom (or group of atoms), which means a participant can dynamically adjust the course and strength of the interaction simply by repositioning their field with respect to the atoms with which they are interacting, until they decide to 'let go'. As an alternative to the Gaussian potential outlined above, we also use spring potentials, a technique used by previous iMD implementations that predate modern virtual reality,[71] which take the following form:

$$\frac{dV_{ext}}{d\mathbf{q}} = 2m_j c(\mathbf{q}_j - \mathbf{g_i}) \qquad \text{Eq (5)}$$

The Gaussian field has the advantage that the maximum force is limited by the Gaussian height, while the spring has no limit. To prevent instability in the molecular system, the maximum force a participant can apply is limited so as not exceed a maximum value.

The Gaussian potential has more flexibility for tuning the strength of the potential, and the fact that it decays to zero at long distances reduces the chance of accidentally exerting a large force on an atom. On the other hand, the spring potential may be more intuitive in some cases, because it allows one to increase the strength of the force by simply increasing the distance. Determining which interactive potential is better for particular applications remains a question for further study in participatory tests. Much of the 'art' of iMD-VR involves understanding how to set the interaction parameters in Eq (4) and Eq (5) so as to enable smooth, stable, and intuitive dynamics for a given dynamics simulation setup, which does not excessively perturb the system. Narupa enables participants to easily modify the value of the scaling parameter $c$ from within VR, tuning the interaction on the fly as they experiment with a given system.

*3.3 Interaction Selection and Force Damping*

The applications discussed in section 4 led us to design new interaction algorithms beyond those described above, in order to facilitate molecular manipulation in more complex systems like biomolecules. In particular, we realized that there were many cases in which it was advantageous to be able to apply a force to an entire subunit of a given molecular system, for example if one wishes to manipulate a portion of a protein's secondary structure and ensure that it remains intact. To address this we implemented a selection interface, shown in **Video 2** (vimeo.com/305459472), which allows a participant to identify a group of atoms which they would like to manipulate (a similar selection interface also enables a participant to choose different renderings for different parts of the molecule). Having specified a particular selection, the participant can then exert an interactive force on the center of mass of the entire subunit, in a fashion that keeps secondary structures intact. Such a method is also extremely useful studying systems linked to protein-ligand binding, for example enabling a researcher to exert an interactive force on an entire ligand. If we let $\mathbf{x}_N$ be the center of mass of the atoms included within a particular selection, and assume that an interactive potential is applied to this group, then the overall force to apply to the atoms, $\mathbf{F}_N$, is calculated as in the single atom case described by Eq (4) and Eq (5), except instead of a single atomic position, the center of mass is used as the center of interaction, effectively substituting $\mathbf{x}_N$ for $\mathbf{q}_j$, and setting $m_j$ to 1, which gives:

$$\mathbf{F}_N = \frac{c}{\sigma^2}(\mathbf{x}_N - \mathbf{g_i})e^{\frac{-\|\mathbf{x}_N - \mathbf{g_i}\|^2}{2\sigma^2}} \qquad \text{Eq (6)}$$

This total force is divided amongst the atoms and applied in a mass-weighted fashion as follows:

$$\frac{dV_{ext}}{d\mathbf{q}} = \frac{1}{N}m_j\mathbf{F}_N \qquad \text{Eq (7)}$$

The resulting interaction allows complex manipulations which for example, can preserve protein secondary structure.

We also realized that, similar to medical surgery, the ability to carry out manipulations on a real-time MD simulation depends critically on a participant's ability to make gentle movements which do not irreversibly perturb too many parts of the system. Interacting with the atomic system by applying bias potentials enables the motion of the system to be integrated as usual. However, in some cases the accumulation of biasing forces on the system can have unintended consequences, as the forces are integrated into the velocities of the atoms of the system. This can make a system challenging to control, because the only way for an atom (or selection thereof) to lose the momentum added by participant manipulation either by: (1) velocity-damping energy transfer through collisions with other parts of the system, (2) velocity dampening and friction from the thermostat, or (3) the participant applying a force in the opposite direction. One strategy which avoids excess momentum build-up during interactive molecular simulations involves performing continuous energy minimization[66] rather than continuously integrating the system dynamics. This strategy works well for small molecular systems and reactions, in which manipulating a single atom and having the system constantly minimize its energy is tractable. Inspired by this strategy, we have developed a hybrid method which uses velocity reinitialization as a way to mitigate the effects of accumulated momentum. Upon interacting with a single atom or group of atoms, the molecular dynamics continues to be integrated as usual, except now the interactive biasing potentials are also being applied. Once the participant stops interacting with the atoms, the atoms involved in the interaction have their velocities randomly drawn from a Maxwell-Boltzmann distribution at a target temperature of $\alpha T$, where $T$ is the target equilibrium temperature of the thermostat, and $\alpha \in (0,1]$ is a scale factor chosen by the participant, which by default is set to a value of 0.5. To maintain stability, velocities are typically reinitialised to a temperature lower than the target equilibrium temperature. This is similar to the Andersen thermostat, except rather than being applied to atoms at random, the velocity re-initialization is specifically targeted at those atoms involved in an interaction. By reinitializing the velocities, any overall momentum in the atoms in a particular direction is removed. Of course, there is a timescale associated with re-equilibration, but applying interactive forces already takes the system out of equilibrium, and the benefit of being able to accurately manipulate groups of atoms which are in an approximately correct ensemble, outweighs this effect.

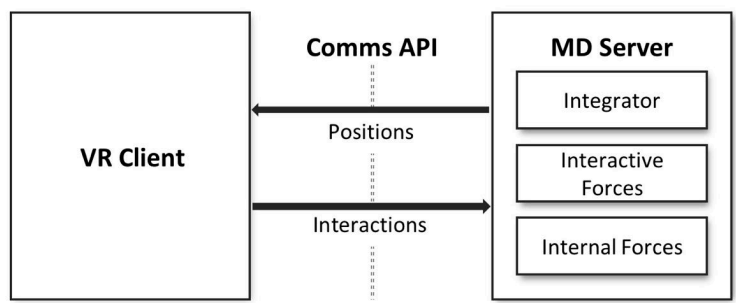

Scheme 1: Broad outline of the Narupa server/client design, and the API that enables communication between the VR client and the MD server. The MD server contains the (thermostatted) integrator, as well as engines for calculating both internal and external forces. The API enables flexibility for connecting different force engines to the VR client.

*3.4 The Narupa Force API*
Simulating the dynamics of a particular molecular system requires an engine to calculate the internal forces. Here we benefit from the fact that our framework has been designed to flexibly communicate with a wide range of force engines via a defined application programming interface (API). As illustrated in Scheme 1, the API functions in a straightforward manner, sending coordinates to a force engine, and receiving forces in return. The idea here is that the force engines which communicate to Narupa can effectively operate as 'black boxes', which simply plugin to Narupa. For example, we have connected our API to the following force engines: an implementation of the MM3 forcefield [44, 72]; the OpenMM molecular dynamics package, which allows access to a range of GPU-accelerated force engines [45]; PLUMED, using the VMD IMD API,[71] which is capable of communicating with a wide range of programs, e.g., GROMACS [48] and LAAMPS;[73] the tight binding density functional theory package DFTB+ [46]; and the semi-empirical quantum chemistry package SCINE.[47] The flexibility of our API enables us to undertake VR-enabled interactive simulations on a wide range of systems, and optionally include either implicit (e.g., continuum) or explicit (e.g., TIP3P water) solvent models. In cases where we model explicit solvent, we do not typically visualize the solvent molecules, in order to maintain clarity and high-quality rendering. Force integration is typically undertaken using a Velocity Verlet integrator, with an Andersen thermostat [74] set to a predefined target temperature. A time step of 1 fs is typical, although we recently implemented the SETTLE and CCMA constrained dynamics algorithms, [75, 76] which enables us to achieve stable dynamics utilizing greater timesteps of up to 2 fs for biomolecular systems. The scientific applications outlined in section 4 benefit from the flexibility of this force plugin architecture. Narupa includes options which enable participants to store trajectories which they generate whilst in VR, for subsequent analysis and post-processing.

*3.5 Narupa renderers*

The flexibility of the Narupa force API enables the simulation of a wide range of molecular systems, and we are consequently working to implement a number of rendering aesthetics. Familiar styles such as ball-and-stick, liquorice and VDW representations are available, as well as a ribbon renderer for protein structures, some of which are shown in **Video 2**. These styles can be applied to any selection layer created in VR, enabling intuitive customization. The visualization settings can then be stored for repeat use or transmitted to other participants to synchronize visualization for shared experiences. High performance rendering of molecular structures in VR is a challenge, requiring a target frame rate of 90 frames per second for each eye, which is further complicated by the requirement for rendering of simulations that are continuously updating from data being received over the network. We are currently working to improve rendering performance, and build additional renderers, such as a secondary structure renderer which can dynamically indicate biomolecular features such as alpha helices and beta sheets. For example, **Video 3** (vimeo.com/315239519) shows a first person perspective of a real-time MD simulation of neuraminidase (PDB 3TI6) displayed using a prototype secondary structure renderer which we will soon add to the main Narupa source distribution. This renderer uses the DSSP algorithm[77] to calculate the hydrogen bonds and secondary structure present in the molecule. This is combined with a cubic Hermite spline passing through the alpha carbon chain of the enzyme to render a continuous 3D chain. The secondary structure assignment is used to color the chain appropriately and to stretch the chain to highlight arrows and helices. The video shows how bits of the secondary structure flicker in and out over the duration of the MD simulation.

*3.6 Narupa examples*

Narupa comes packaged with a number of stable examples, which participants can inspect in order to guide them in setting up their own interactive simulations. At present, the following examples are packaged with Narupa:

- Two $C_{60}$ buckyballs at 300K simulated with a timestep of 1fs. This is the usual introductory simulation for familiarizing users with the iMD-VR environment.
- A carbon nanotube and methane molecule simulated at 200K with a 0.5fs timestep. The 'task' here is to pass the methane molecule through the nanotube.
- A short helicene fragment at 300K and with a 1fs timestep, which users can manipulate to switch between conformations characterized by either a left or right-handed screw sense.
- A 17-ALA helical peptide chain at 300K and with a 2fs timestep simulated with the Amber99SB forcefield, used to demonstrate the ability to tie a molecular knot. This simulation requires the OpenMM package.
- The enzyme H7N9 Neuraminidase and the drug Oseltamivir, to demonstrate drug unbinding and rebinding, with the protein using the Amber03 force field, and the drug force field parameterized using GAFF. This simulation is run using an Andersen thermostat at 300K, with a Verlet integrator with timestep 0.5fs. This simulation also requires the OpenMM package.
- The smallest known knotted protein MJ0366 in its native state, to illustrate the utility of 3D visualization, simulated with the Amber03 forcefield using an Andersen thermostat at 300K with a Verlet integrator with timestep 0.5fs. This simulation requires the OpenMM package.

Unless otherwise specified, all simulations use the Berendsen thermostat and the velocity Verlet integrator. The three hydrocarbon simulations all use the MM3 force field. In the near future, we will be adding a number of additional examples to the open-source repository (e.g., those outlined in section 4). For the smaller simulations, good performance and fluid interactivity can be achieved by running the force engine server and VR render client on the same machine. However, for the larger simulations (e.g., H7N9 Neuraminidase, MJ0366, or the quantum chemical systems described in section 4), achieving good performance & fluid interactivity often requires running the force engine on one machine and the VR render client on another, with communication over a fast local network.

## 4. Scientific Research Applications

*4.1 Measuring Task Completion Times*

As we have come to demonstrate this framework more extensively, we have often encountered the question: 'does this provide any research benefit?' To answer this question, we published recent work aimed at quantitatively evaluating the extent to which our framework accelerated some typical molecular simulation tasks. In a series of controlled HCI studies, we tasked participants with a range of molecular manipulation goals: (1) threading methane through a nanotube; (2) changing screw-sense of a helicene molecule (from left to right handed); and (3) tying a protein knot. These tasks were selected because each requires a complicated 3D dynamical choreography, and also because they represent distinct classes of dynamical change which are important across natural and engineered nano-systems. For example, the $CH_4$/nanotube task provides an analogue for transport dynamics across nano-pores of the sort which are ubiquitous across bio/materials chemistry.[78] The helicene task provides an example of induced changes in molecular helicity, which synthetic biologists have explored as a strategy to transmit chemical messages.[79,80] Molecular knots have been successfully designed in recent synthetic work,[81] and have also been observed to occur in protein structures, where they have been associated with neurodegenerative diseases.[82] The results, shown in Figure 2, quantitatively demonstrate that participants within the iMD-VR environment can complete molecular modelling tasks more quickly than they can using conventional interfaces like a mouse or a touchscreen, especially for molecular pathways and structural transitions whose conformational choreographies are intrinsically 3-dimensional.

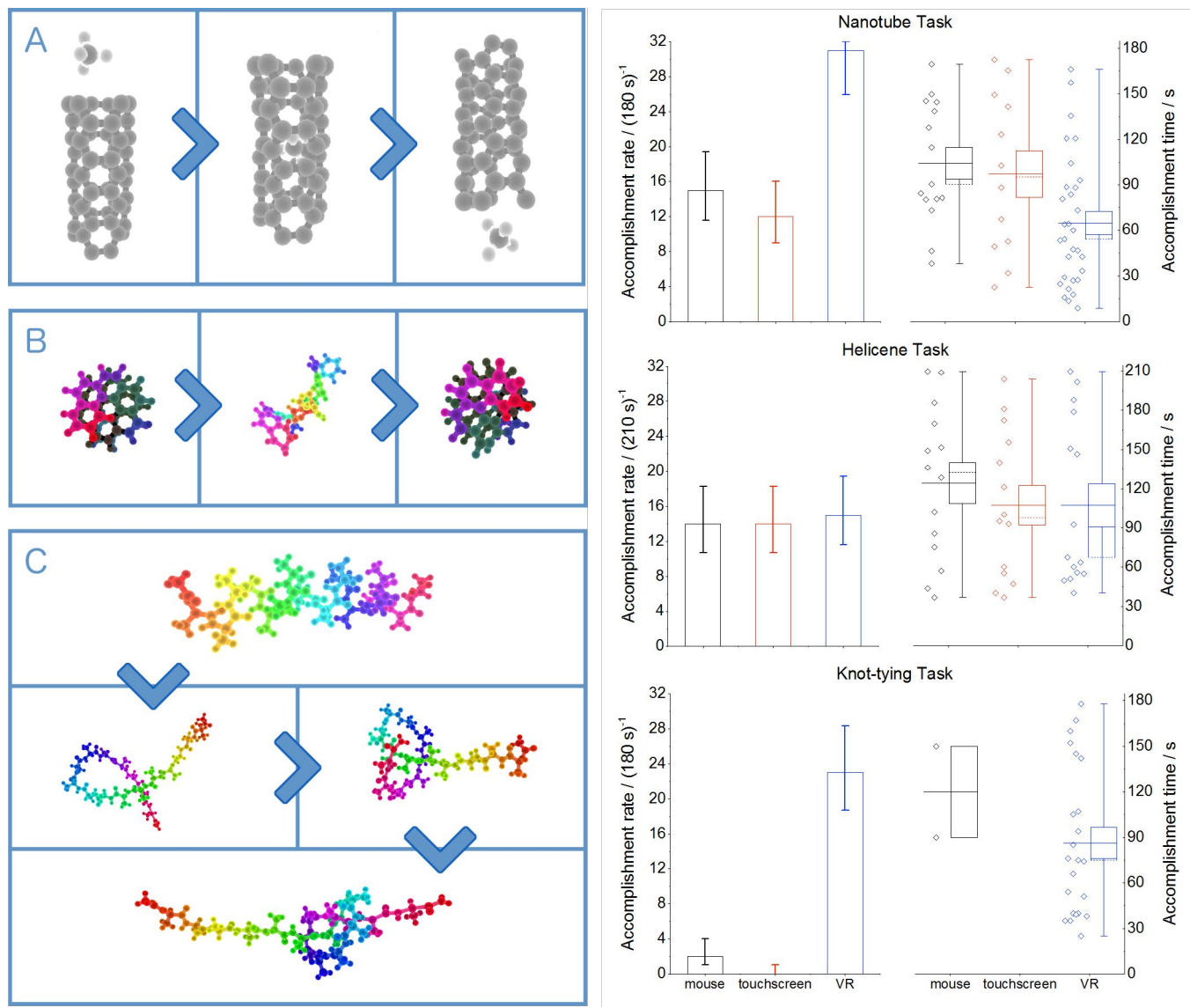

Figure 2: left hand panel shows interactive molecular simulation tasks used as application tests: (A) threading $CH_4$ through a nanotube; (B) changing the screw-sense of a helicene molecule; and (C) tying a knot in a polypeptide (17-ALA). Colors selected in this figure are chosen for the sake of clarity. The right hand panel shows the user study results corresponding to each task, including accomplishment rates for each task (n = 32 for all tasks), with Poisson error estimates, and the distribution of task accomplishment times. In the box-and-whisker plots, whiskers indicate the data range and the box indicates the standard error of the distribution. The mean is shown as a solid line, and the median as a dashed line.

For tasks A and C, Figure 2 indicates that iMD-VR provides a clear acceleration benefit compared to the other platforms, and also that - the more inherently 3D the task, the greater the benefit. The knot-tying task results (Figure 2C) are the most dramatic. A task like knot-tying, which is so intrinsically 3D, is very difficult to accomplish outside of VR. For the nanotube task (Figure 2A), the accomplishment rates, mean time, and median time in iMD-VR are a factor of approximately 2x – 3x faster than on other platforms. At first glance, the helicene task (Figure 2B) is a case in which iMD-VR appears to provide little significant rate enhancement compared to other platforms. Observation of the study participants show that this is because changes in helicene screw-sense are most efficiently accomplished using a simple 2D circular motion, as shown in **Video 1**. Essentially, the 2D limitations of the mouse and touchscreen constrain the participant to carrying out a motion which is well suited to inducing changes in molecular screw-sense, so that iMD-VR provides little additional benefit. Closer inspection of the helicene time distributions shows that iMD-VR does afford some advantage: the median time required to change molecular screw-sense in iMD-VR is 30–40% less than the median time required on a touchscreen or using a keyboard/mouse.

*Reassuringly, we found zero instances where users experienced VR-sickness during the experiments carried out to gather the data in Figure 2*. To date, thousands of people have volunteered to experience our system, and very few (less than ten) instances have arisen where participants report any form of sickness – a very small probability. This is an important point, because there is a widespread misconception that a VR experience necessarily entails some form of motion-related illness. This is not in fact the case. The causes of VR sickness are well understood by workers in human computer interaction and psychology. One of the most common causes of VR sickness arises from inconsistency between the visual information arriving to the brain and the information arriving for processing by the vestibular and proprioceptive system. For example, a sure-fire way to induce VR sickness is by simulating motion within the VR headset whilst a participant is stationary. In such a case, the brain's visual system is presented cues suggesting motion, at odds with the cues to the vestibular and proprioceptive systems, which are not experiencing motion. This perceptual disconnect leads to sickness in significant fractions of people (including several of the authors on this article!). For high-performance scientific applications like those being discussed herein, VR sickness can sometimes arise as a result of computational bottlenecks which cause the system to 'lag'. In such cases, it is often possible to improve system performance through detailed optimizations, or at least to define the operational performance limits of the system which avoid participants experiencing illness. The important point is this: *high-end commodity VR enables designers to avoid experiences which lead to illness*. In the vast majority of cases, the origins of VR sickness are well-understood, and neither designers nor participants should settle for VR experiences which induce illness.

*4.2 Using iMD-VR to train neural networks to learn reactive PESs*

The first iMD-VR simulation we ever ran (in Sept 2016) was $OH + CH_4$, using a multi-state EVB force field[83, 84] which we designed to simulate $OH + CH_4 \rightarrow CH_3 + H_2O$. Tests run on a number of participants indicated that our system provided insight into several subtle nuances characteristic of dynamical systems. For example, participants reported that: (1) when using their handheld force 'tweezers' to manipulate the $CH_4$ Carbon, they could detect the vibrational wobbliness of the much lighter attached Hydrogens; (2) as they brought $OH + CH_4$ into close proximity, they could feel the non-local electrostatic repulsions which arise; and (3) when releasing the OH or $CH_4$ molecules with translational kinetic energy, they could see the resulting translational and vibrational damping as a consequence of the thermostat. Successfully undertaking a reaction to make $CH_3 + H_2O$ required that participants place sufficient kinetic energy into the relative translational motion of the reactants to get over the barrier, and also that they guide $OH + CH_4$ into an orientation which could overcome the randomizing influence of entropy and enables hydrogen abstraction. These early experiments, illustrated in **Video 4** (vimeo.com/315218999), provided the first indication that our iMD-VR system was sufficiently intuitive and afforded adequate control for simulation tasks to be successfully undertaken in a way which was reproducible by a wide cross-section of participants.

In general, the simulation of chemical reactions requires quantum mechanical approaches, which are computationally expensive compared to molecular mechanics, and limit the size of simulation that can be performed. For example, parallelized semi-empirical methods enable us to explore systems with 100 – 200 atoms at interactive latencies. Using machine learning, it is possible to train models which are faster than quantum mechanical methods, and which reproduce quantum mechanical energy surfaces. To enable the interactive simulation of even larger systems, we been exploring using Narupa as an iMD-VR strategy for rapidly sampling chemical space and building up data sets which can then be used to train machine learning algorithms in order to learn potential energy functions.

The rise of machine learning has resulted in an interesting paradigm shift which sees increasing value being placed on *data curation*—that is, data size, quality, bias, format, and coverage. Data-related issues are becoming as important and time-consuming as the algorithmic methods used to process and learn from the data, and iMD-VR provides an efficient strategy whereby human experts can curate data which can then be used to train machines.

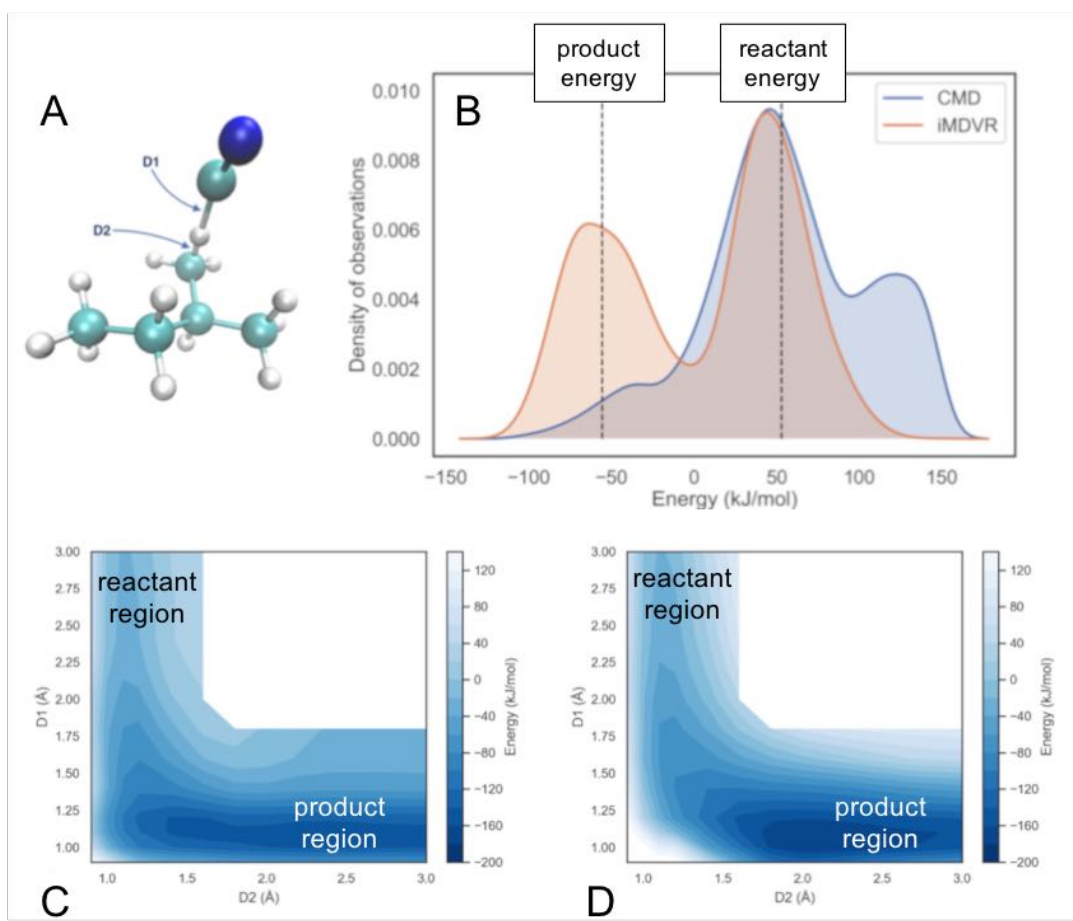

Figure 3: (A) shows the bond distances defining the Hydrogen abstraction reaction coordinate. (B) shows the Kernel density estimate of the configurational energies sampled using the iMD-VR approach (orange) and the constrained MD approach (blue). The dotted lines show the average energy of the sampled reactant and product geometries. Panels (C) and (D) show the PESs predicted by Neural nets trained using geometries sampled using iMD-VR and CMD, respectively.

Curating data sets of molecular geometries and their corresponding *ab initio* energies, which can then be fit using functions that yield efficient energies and gradients on the PES, represents a longstanding research problem in computational molecular physics. In recent work, we have shown that exploration of molecular configuration space by human participants using iMD-VR to steer 'on-the-fly' *ab initio* MD can be used to generate molecular geometries for training GPU-accelerated neural networks (NN) to learn reactive potential energy surfaces (PESs).[13] **Video 5** (vimeo.com/311438872) shows our first application using this strategy, focused on hydrogen abstraction reactions of CN radical + isopentane using real-time semi-empirical quantum chemistry through a plugin to the SCINE Sparrow package developed by Reiher and co-workers[85-87] (scine.ethz.ch), which includes implementations of tight-binding engines like DFTB alongside a suite of other semi-empirical methods.[47] To obtain the results described herein, we have utilized the SCINE Sparrow implementation of PM6, with the default set of parameters. Using real-time PM6 in VR, graduate student Silvia Amabilino spent approximately one hour in iMD-VR, using it to sample a wide range of H-abstraction pathways at primary, secondary, and tertiary sites on isopentane. Figure 3 shows an illustrative example abstraction of a primary Hydrogen, and compares the PESs predicted by NNs trained using data obtained from iMD-VR versus NNs trained using a more traditional method, namely molecular dynamics (MD) constrained to sample a predefined grid of points along those coordinates which define hydrogen abstraction reactions (shown as D1 and D2 in Figure 3A). Figure 3B shows the density of points sampled for each method as a function of energy; the bimodal structure of the iMD-VR curve reflects sampling in the product and reactant minima, indicating that user-sampled structures obtained with the quantum chemical iMD-VR machinery enable excellent sampling in the vicinity of the minimum energy path (MEP). The data in Figure 3B provides strong evidence that iMD-participants can generate data which is clustered along the MEP, and therefore not too far from an equilibrium ensemble. Constrained MD data (CMD), in comparison, did less well in sampling along the MEP, but enabled sampling of high-energy 'off-path' structures. Figure 3C shows the predictions of NNs trained using iMD-VR data and Figure 3D shows the predictions of NNs trained using the constrained MD data. Both reproduce important qualitative features of the reactive PESs such as a low and early barrier to abstraction. The NN trained on the iMD-VR data does very well predicting energies which are close to the MEP, but less well predicting energies for 'off-path' structures, whereas the CMD data does better predicting high-energy 'off-path' structures.

More broadly, we are developing an API plugin which will enable communication between Narupa and the QML quantum machine learning package initiated by von Lilienfeld and co-workers,[88] so that QML can be used as a force engine for iMD-VR. QML has available a wide range of machine learned models to provide forces and energies trained on high-level quantum data. Kernel-based models like those available in QML can be used to describe molecular potential energy surfaces with spectroscopic accuracy, using only a very limited amount of training data. Such models are inherently fast, with $O(N)$ scaling if used with appropriate cut-offs.

*4.3 Reaction discovery using 'on-the-fly' ab initio dynamics*

A particularly prevalent problem in the chemical sciences involves mapping complex networks of reactions in order to predict how a particular system (e.g., the gas mixture in a combustion engine, [89] or a complex catalytic cycle [90]) evolves in time. Devising automated methods for discovering important reactions and transformations characterizing a given chemical system is an area that has attracted significant interest in recent years, with a number of strategies proposed to tackle the problem. [89, 91-96] Building on a number of recent examples where scientific problems have been 'gamified', [97-99] we have been using Narupa to investigate the extent to which human insight might be harnessed to accelerate mechanism discovery and understand how human search differs from machine search.

Video 6 (vimeo.com/312963823) shows a participant's first-person perspective as they manipulate a real-time simulation using a quantum mechanical force engine to 'discover' chemical reactions in the OH + propyne system. Figure 4 shows preliminary data obtained from a participant group of 21 University of Bristol undergraduate students, each of whom were given five minutes using iMD-VR in Narupa to discover reactions in this way. In our preliminary tests, the students were given a very simple instruction to 'discover' as many different reactions as they could. Forces in these simulations were obtained thorough an interface with the semi-empirical Scine code using the PM6 level of theory. [47] Figure 4 shows a comparison of these preliminary participant-generated results with those obtained from the ChemDyME automated reaction mechanism generator (github.com/RobinShannon/ChemDyME) using the same level of theory.[95] In Figure 4 each node in the network diagram represents a different molecular configuration, all originating from the green OH + propyne node. Figure 4 shows that iMD-VR participants and ChemDyME initially found many of the same reactions, represented by the red nodes. The reactions sampled in iMD-VR (blue) and by ChemDyME (orange) then diverge, characterized by two very different search strategies. ChemDyME sampling covers a smaller number of reactions with lots of dead end nodes, whereas human guided VR-sampling identifies many more channels, with significant inter-conversion between nodes. Preliminary analysis indicates that human guided iMD-VR searches were particularly adept at finding association and dissociation processes – e.g., involving high energy association and dissociations of a single species into 2 or more fragments. In this instance, ChemDyME appears better at finding isomerization barriers.

We devised a preliminary 'scoring function' for comparing the performance of the respective search strategies. The scoring function awarded points for finding new pathways. In an attempt to incentivize iMD-VR users to discover lower energy pathways, less points were awarded for higher energy pathways. Figure 5 shows this scoring function applied to the iMD-VR and the ChemDyME results as a function of the number of timesteps. iMD-VR participants were extremely effective at finding a large number of high scoring bimolecular channels. Compared to ChemDyME, our preliminary scoring function implementation appears to incentivize human experts to find more channels overall, but to miss lower energy channels. Moving forward, we plan to investigate the extent to which different scoring mechanisms in conjunction with auditory feedback might influence search strategies. We are particularly interested in understanding the difference in human vs. computer search strategies, and understanding whether human search techniques might be used to devise new kinds of automated search algorithms. [99]

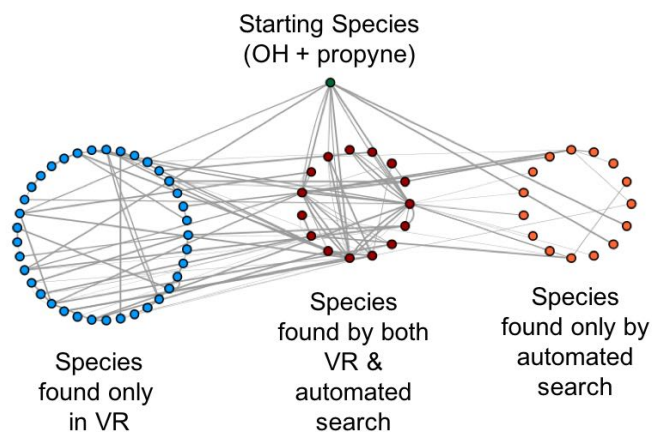

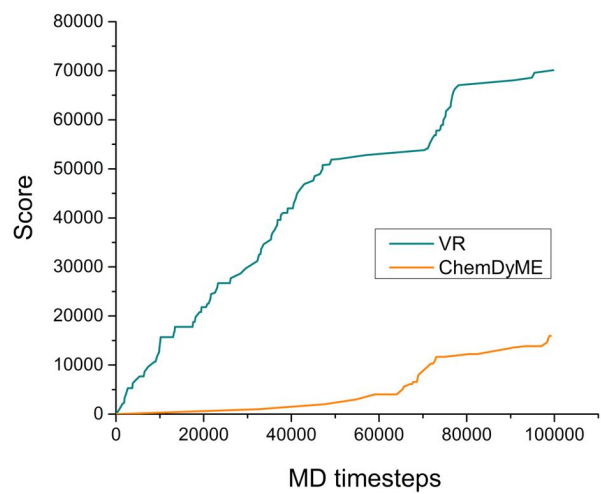

Figure 4: Comparison of reactions (edges) and species (nodes) found by humans in iMD-VR vs. those from the automated ChemDyME software. The starting node (OH + propyne) is green, those chemical species found in both iMD-VR and ChemDyME are in red, those found in iMD-VR only are in blue and those found from ChemDyME only are in orange.

Figure 5: The time-dependent score as reactions are discovered, using both iMD-VR and the automated ChemDyME software

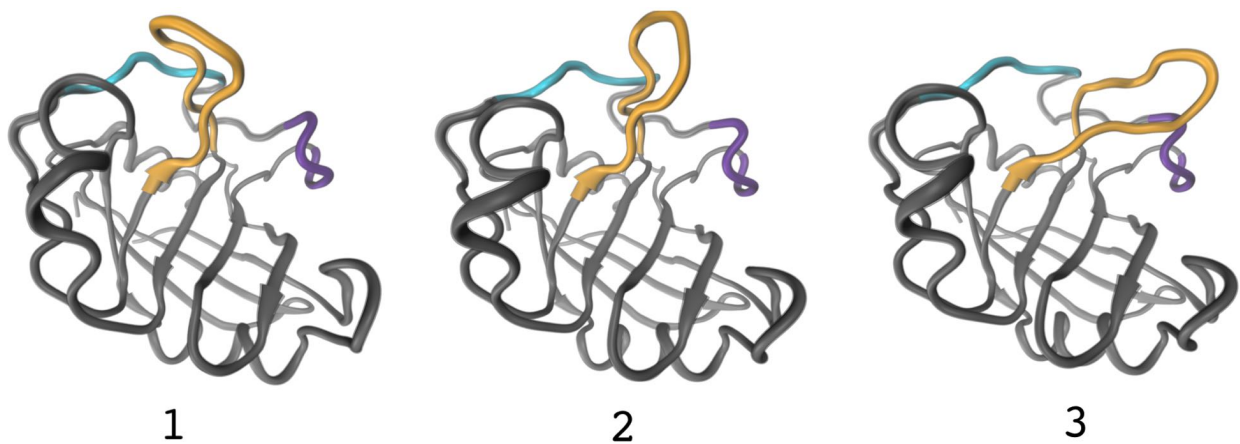

Figure 6: Conformational snapshots of CypA secondary structure sampled during a single representative iMD-VR trajectory, where the participant was specifically exploring gating motion of the 100s loop (colored orange). Conformational snapshot 1 shows CypA in its native state, with the 70s loop (formed by residues 65 – 75) colored purple, and residues 80 – 90 colored cyan. Conformational snapshot 3 shows the 100s loop is in close proximity to the 70s loop. Conformational snapshot 2 shows an intermediate conformation, where the 100s loop lies between the 70s loop, and residues 65-75.

*4.4 Measuring Task Reversibility in Complex Systems*

If iMD-VR is to evolve into a sophisticated tool for carrying out detailed atomic manipulations on systems which are larger and more complicated than those discussed above, then a critical question is the extent to which complex structural manipulations – e.g., in a biomolecule – are in fact reversible. The level of reversibility is an indicator of: (1) the level of control which a participant has over the systems they are investigating, and (2) the extent to which the iMD-participant is able to maintain the system in an ensemble which is not too far from equilibrium. In a first attempt to try and evaluate this, we have been looking at loop motions in the well-studied protein cyclophilin A (CypA), where there is evidence that large-scale collective motions take place. [100, 101] Here we highlight some preliminary results which we have obtained during studies of the so-called '100s' loop in CypA (formed from residues 100-110), which undergoes a gating motion shown in Figure 6. The representative configurations shown in Figure 6 come from an interactive trajectory generated using iMD-VR, whichis shown in **Video 7** (vimeo.com/306778545). Snapshot 1 of Figure 6 shows the CypA native state, in which the 100s loop (highlighted in orange) is in close proximity to residues 80-90 (colored cyan). Snapshot 2 shows a conformational state in which the loop has begun to move away from residues 80–90, and toward the purple 70s loop (composed of residues 65-75). In Snapshot 3, the 100s loop is in very close proximity to the 70s loop.

Starting from the native state shown in Figure 6A, we generated three different iMD-VR trajectories, with our intent being to move the loop away from its native structure, and then back again, following a similar progression as shown in Figure 6. Figure 7 shows the fraction of native contacts along each of the three iMD-VR trajectories A, B, and C. It shows that two of the trajectories (B and C) make excursions away from the native state before returning towards it. The trajectory coloured in orange (A), however, does not returns close to its original native state structure over the duration of the iMD-VR run. Closer inspection of trajectory A shows that the 100s loop did indeed move towards the 70s loop (as intended), but that upon returning back toward the native state, the loop contained too much momentum. As a result, it irreversibly distorted the structure. The right hand panels of Figure 7 tell a similar story as the left hand panel, but using a slightly different representation: here we show the time-dependence of the trajectories in the space of the first two principle components of the heavy atom contacts. In both Figures 6 and 7, iMD-VR trajectory C (colored green) is particularly noteworthy: it shows that a careful participant can return the loop to a configuration which is very close to the native state, with 0.996 of all native contacts restored. Whilst the sample size here is relatively small, this is nonetheless an encouraging result: it shows that, if molecular manipulation is carried out with requisite attention to detail, then it is possible to perform subtle, reversible manipulations of the protein structure from within an iMD-VR environment.

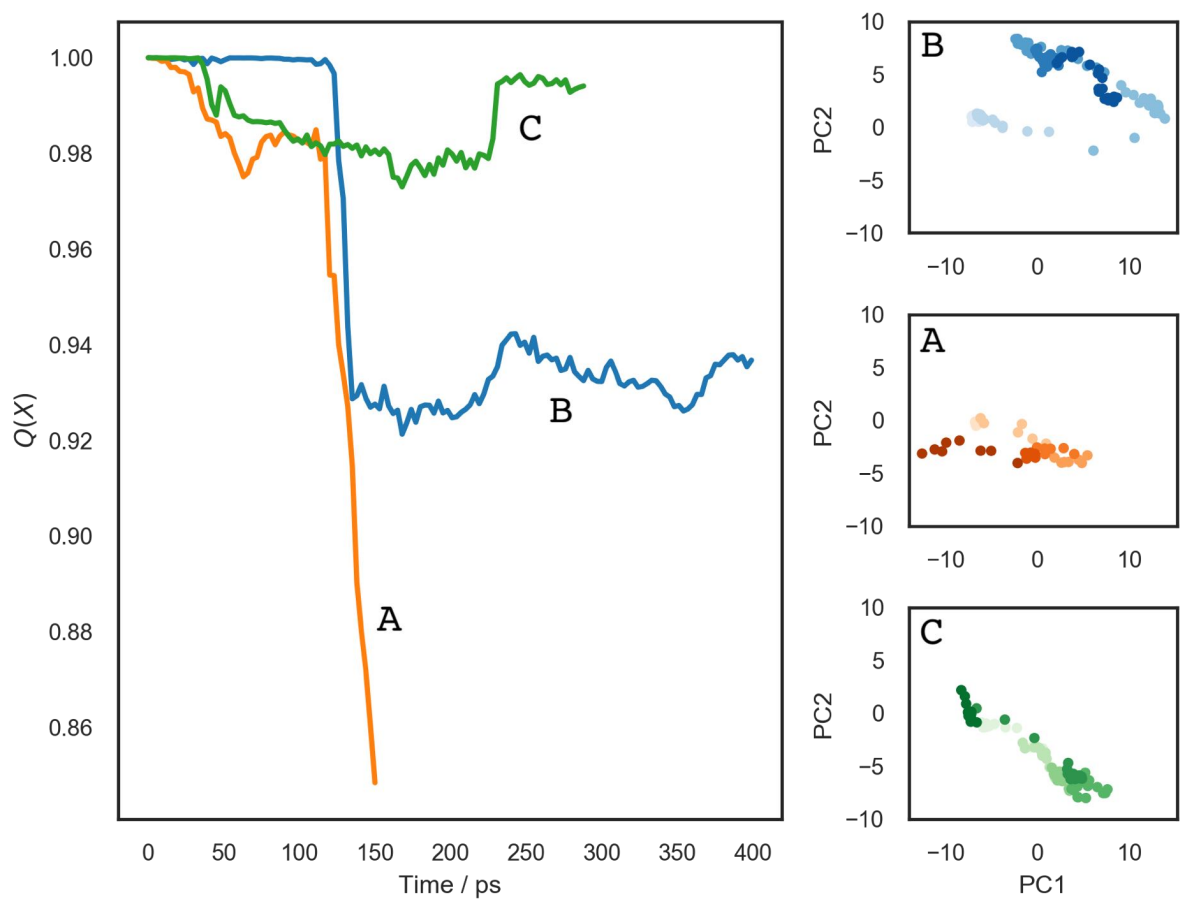

Figure 7: The left hand panel shows fraction of native contacts over the course of the three iMD-VR trajectories (labelled here as A, B, and C) in which the 100s loop motion of CypA was explored; right hand panel shows the corresponding iMD-VR trajectories projected onto the first two principal components obtained from PCA using the all heavy-atom contact distances as the features. Trajectories are coloured from light to dark to indicate the passage of time.

*4.5 Protein-ligand binding*

The preliminary results outlined above suggest that complex biomolecular manipulations using interactive molecular dynamics in iMD-VR are in fact reversible. With this knowledge in hand, we have been exploring additional biomolecular application domains where iMD-VR might be used to provide insight into biomolecular structure, function, and dynamics. One specific domain where we have been concentrating our efforts involves the use of iMD-VR to undertake flexible docking of small molecule ligands to protein structures, as illustrated in Figure 8. Broadly speaking, the discovery of molecular binding poses using interactive molecular dynamics amounts to a four-dimensional puzzle in which correct solutions are found by moving, rotating, and fitting a ligand into a protein binding pocket. Whilst there are increasing efforts aimed at using molecular dynamics to examine protein-ligand binding, [102-105] an iMD-VR approach focuses on providing experts with a straightforward and intuitive means for expressing their molecular design insight to evaluate potential drug designs and corresponding binding hypotheses. Using Narupa, we have been exploring the extent to which human design insight and spatial reasoning can be used to guide binding hypotheses, discover potential binding poses, and generate dynamical binding pathways for analyzing binding kinetics and mechanisms. Resolving the kinetic mechanisms of the ligand-protein association process has increasingly been recognized to provide additional insight into safe and differentiated responses of candidate therapeutics.[103]

For example, **Video 8** ([vimeo.com/274862765](vimeo.com/274862765)) shows interactive binding experiments which we undertook to dock the benzamidine ligand to the trypsin protein using the Narupa-OpenMM interface. Specifically, the video was generated beginning from the benzamidine-trypsin complex (PDB:1S0R), which we parameterized using GAFF and the Amber14SB forcefield, treating solvent effects using an OBC2 generalized Born implicit solvent model. In order to ensure the iMD-VR user didn't accidentally perturb the tertiary structure of trypsin, a restraint was applied to the protein backbone; however, subsequent tests without restraints have since shown that careful iMD-VR users are able to undertake drug unbinding and rebinding without perturbing protein tertiary structure. The movie shows benzamidine being interactively guided out of the trypsin binding pocket, and then re-docked. Our preliminary results, established through tests carried out in collaboration with participants at a recent UK CCP-BioSim workshop, indicate that participants, starting from a state where benzamidine was undocked, were then able to identify the trypsin binding pocket and subsequently generated a dynamical pathway which established a bound pose. These preliminary results provide evidence that it is indeed possible to accelerate protein-ligand binding rare events, and also that the Narupa toolset furnishes sufficient control for this class of rare events to be reversible, consistent with the conclusions of the previous section. Combined, these results suggest that the spatial cognition of a trained biochemist can furnish insight into protein-ligand binding events, in order to quickly explore a wide range of thermodynamic states and kinetic pathways. By combining molecular insight and spatial reasoning, participants were able to manipulate the benzamidine in a fashion that allowed the primarily electrostatic binding forces to be overcome, and then reestablished. Preliminary results which we have undertaken to investigate the docking of oseltamivir (commercially known as Tamiflu) to the H7N9 strain of avian flu neuraminidase are similarly encouraging, and indicate that docking can be achieved even in a system where the docking dynamics are more complicated, where unbinding and rebinding require the opening and closing of a protein loop, as shown in Figure 8 and **Video 9** (vimeo.com/296300796).

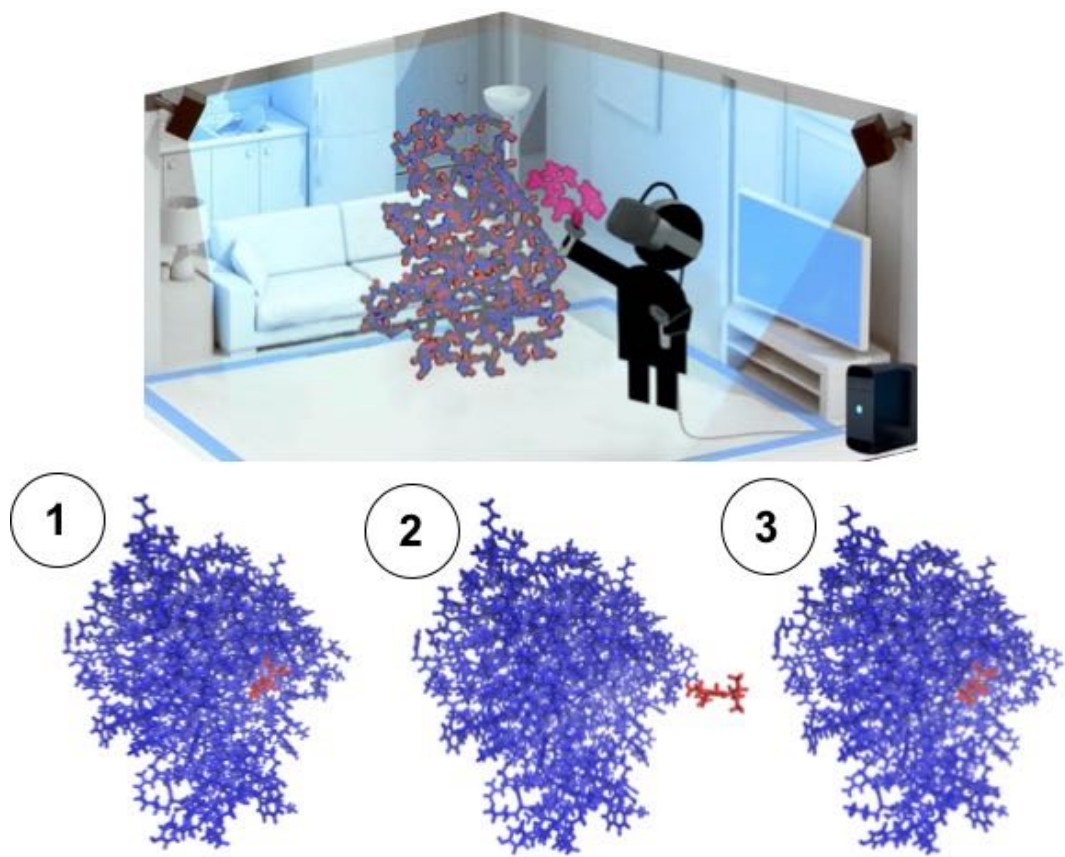

Figure 8: Top panel illustrates a researcher using iMD-VR investigate pathways for binding and unbinding a ligand (magenta) to a protein. Snapshots 1 – 3 in the bottom panel show three of the geometries generated as a researcher utilizes Narupa to interactively manipulate a real-time Amber MD simulation of H7N9 neuraminidase and oseltamivir, exploring dynamical pathways for unbinding and rebinding oseltamivir. Conformational snapshot 1 shows a structure near the beginning of the iMD-VR session, where oseltamivir is bound to neuraminidase; conformational snapshot 2 shows oseltamivir after the iMD-VR researcher has undocked it from neuraminidase; and conformational snapshot 3 shows the final pose once the iMD-VR researcher has re-docked oseltamivir to neuraminidase.

*4.6 Molecular Transport in Zeolites*

We have also been applying the Narupa iMD-VR framework to understand the transport of small molecules through periodic solid-state materials like zeolites [106] and metal-organic frameworks (MOFs). [107] Compared to protein structures of the sort discussed above, nanoporous materials like these can we constructed from a number of different elements, and are often characterized by a similarly wide range of distinct bonding patterns. Whereas the important interactions governing protein-ligand type interactions tend to occur relatively near the surface of protein structure, the same is not true for small molecule transport in structures like zeolites. Small molecule transport in structures like these tends to occur in channels which are buried in the interior, and which have a complex branched structure, which can lead to transport which involves non-intuitive directionalities. Such structures are particularly important for industrial applications owing to the fact that they are able to accommodate small molecules like hydrocarbons, facilitating both transport [108, 109] and catalysis. For example, within the petrochemical industry, these sorts of materials have essential functions as catalysts for processes like hydroxylation, alkylation, and epoxidation, [106] where they operate at much higher temperatures and pressures than typical biocatalysts.

The fact that such materials typically find application under more extreme conditions means that studying them in iMD-VR requires a force regime which is quite distinct from those which are typically used in our biomolecular studies. It also means that these structures are more robust to the formation of local 'hotspots' of the sort that can sometimes arise in iMD-VR applications. Figure 9 and **video 10** (vimeo.com/312957045) shows a ZSM-5 zeolite structure which we have recently begun to study using Narupa, in order to better understand the transport kinetics of 2-methyl-hexane. In order to study this particular system, we connected the Narupa API to PLUMED, which enables communication with a wide variety of force engines including DL_POLY,[49] from which we obtained forces. The PLUMED interface allows us to retain the full flexibility of the DL_POLY program and run simulations using any of its internal MD parameters and methods. The system shown in Figure 9 is comprised of 288 zeolite atoms, and was set up to be fully periodic with a vacuum gap of 10 Å along the Z axis. The Langevin thermostat was set to 648K with a friction coefficient of 5 ps$^{-1}$. The left hand panel of Figure 9 shows a first-person participant's perspective as they manipulate the zeolite; the right-hand panel shows a partially extracted hydrocarbon in a van der Waals representation. As the video shows, iMD-VR enables one to perform detailed inspection of the zeolite microstructure, interact with substrates in order to navigate them within the channels, and test a range of pathways in order to understand the mechanism and kinetics for adsorption, desorption, and transport. In our preliminary studies on small-molecule transport through zeolite frameworks, we have found that the ability to manipulate and deform the channel has enabled us to better understand how the channel structure and its corresponding flexibility impacts on the hydrocarbon transport dynamics.[110]

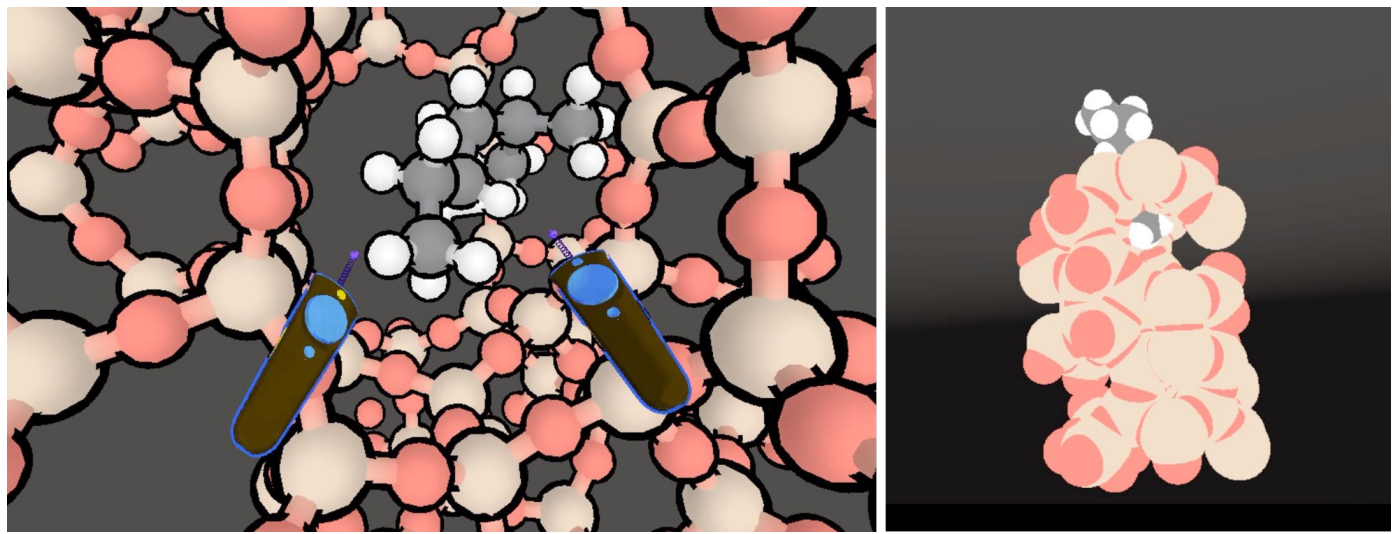

Figure 9: ZMS-5 2ME-hexane structure. The left hand panel shows interaction with a methyl group using the NarupaXR controllers, in order to manipulate the hydrocarbon position. The right hand panel shows the van-der Waals radius representation for the same structure where the hydrocarbon has been partially extracted from the structure.

## 5. Ongoing HCI Research

*5.1 Sound as a real-time data channel*

In a typical iMD-VR simulation, a massive quantity of data is available for real-time streaming, much more than a participant can easily synthesize and understand on-the-fly. One of the challenges of using the Fig. 1 framework arises from the fact that there are a range of important interactions which researchers often care about during the course of a simulation, but which can be difficult to render 'on-the-fly' as a result of *visual congestion*. There is simply not enough screen space nor graphical processing power to render all of the interactions a researcher may care about as they use iMD-VR to explore a given system. One potential solution would be to create multiple real-time visual displays (e.g., plots) within the iMD-VR environment; however, this risks splitting the participant's attention away from the molecular system which they are studying. Furthermore, many of the features which are important during the course of an iMD-VR simulation – nonbonded energies, potential energy, steric clashes, or the instantaneous values of complex collective variable values – are difficult to represent visually. Data sonification – where real-time data streams are converted into audio channels – offers a potential solution to these problems: it allows for the possibility that auditory channels can be used to convey important information, thereby reducing the quantity of onscreen rendering congestion, and it offers ways of tracking real-time data streams that are difficult to render visually.

Apart from a few examples,[111, 112,113] data sonification represents relatively unexplored territory for the molecular sciences. However, research in human perception has shown that audio information can have an important impact on visual perception; in some cases audio perception even overrides visual perception.[114] Moreover, our auditory system has a significantly faster time resolution and a larger dynamic range than our visual system. Audio can also be spatialized audio (e.g., we can 'hear *where* things are coming from', even when the sound source is not directly pointed at our ears). These features of audio make it an indispensable information channel in video gaming and flight simulators, and an interesting data channel to explore in an immersive iMD-VR environment. Inspired by studies showing that multisensory integration of vision with sound improves our ability to accurately process information,[3] we have begun to explore how sound might be utilized to augment structural visual information in the molecular sciences.[6]

The underutilization of sound in the molecular sciences arises in part because methods for auditory rendering are less well defined than their visual counterparts. Depicting an atom as a sphere and a bond as a stick is an arbitrary decision, but is intelligible owing to the fact that both an atom and a sphere are spatially delimited. Defining such a clearly delimited object in the audio realm is not straightforward, neither spatially nor compositionally: it is difficult to imagine what constitutes an 'atomistic' object in a piece of audio design or music. Our preliminary work suggests that sound is best utilized for representing non-local and dynamic properties of the sort which are important in molecular science – e.g. potential energy, free energy, non-bonded energy, electrostatic energy, local temperature, strain energy, etc. Owing to their non-locality, properties like these are difficult to represent using conventional visual rendering strategies. Sound, however, offers an excellent means for representing these things. Moreover, their rapid dynamical fluctuations have a better chance of being detected by the auditory system, given its faster temporal resolution compared to the visual system. **Video 11** (vimeo.com/312994336) shows a real-time iMD-VR simulation of 17-ALA peptide in which the potential energy can be heard as a real-time audio stream. As the participant manipulates the peptide, taking it from its native folded state to a high-energy knotted state which is kinetically trapped, the sound dynamically changes. Eventually, the participant unties the peptide knot, and the protein relaxes to a lower energy state, which is reflected in the sound. Interactively rendering the real-time potential energy of a molecule poses a serious challenge for visual display methods, owing to the fact that the potential energy of a molecule is a non-local descriptor which depends on the entire coordinate vector **q**. However, audio effectively captures changes in the energy. We are currently undertaking user tests to evaluate the extent to which audio improves accomplishment for tasks like drug-ligand binding.

*5.2 VR Gloves: Beyond controller-based interaction*

Several participants who have experienced Narupa have remarked that the controllers act as a barrier in their ability to feel the dynamics of the simulated molecular systems.[9, 69] As a result of these comments, we have been pursuing another avenue of research which involves the ability to reach out and 'directly touch' real-time molecular simulations in iMD-VR – i.e., without being mediated by a wireless VR controller like that wielded by the participants shown in the Figure 1 schematic. We have been experimenting with a wide range of VR-compatible glove technologies. For example, technologies like the Noitom Hi5 glove and the Manus VR glove, which are equipped with 9 degree of freedom (DOF) inertial movement units (IMUs) and several finger mounted bend sensors, are able to interactively track the relative position of the hand. To get positional tracking within a VR system like that shown in Figure 1, these gloves can be combined with a wrist-mounted HTC Vive Tracker. The commercially available Noitom Hi5 and Manus VR gloves are primarily designed for gestural tracking in order to distinguish amongst different hand poses for application in motion capture studios (e.g., identifying a 'thumbs-up' gesture versus or a Vulcan 'live long and prosper' gesture). However, these gloves are relatively expensive, and we have found that their performance is not particularly well suited to the kinds of tasks that a molecular scientist might want to carry out in VR. For example, in an iMD-VR simulation, the ability to accurately distinguish between hand poses is far less important than the ability to accurately detect when a molecular scientists is reaching out to 'grasp' a particular atom (or selection of atoms) between their thumb and their forefinger.

We have made progress in designing our own data gloves,[50] a prototype of which is shown in Figure 10. By sewing modern conductive fabrics into the glove, we can detect when a participant closes one of two circuits, either by making a pinching motion between (a) their thumb and index finger, or (b) their thumb and forefinger. The absolute position of the hand is obtained from mounting an HTC Vive Tracker on the back of the glove. Our preliminary results, obtained from a small set of user studies carried out in our own laboratory, suggest that participants find the molecular and atomic interaction afforded by this glove extremely intuitive. Compared to the standard HTC Vive handheld controllers (shown in Figure 1), participants have indicated that they prefer the direct sense of 'touching' virtual simulations afforded by this glove. Moreover, for accomplishing a range of molecular manipulation tasks, we have found that the Figure 10 glove design results in more stable iMD-VR experiences than the much more expensive Manus and Noitom gloves. **Video 12** (vimeo.com/305823646) shows the perspective of participant who is wearing these gloves as they undertake a protein knot-tying task. To 'touch' an atom and exert a force on it, the participant simply reaches out to the atom they wish to touch, and brings together their thumb and forefinger, as they would do if they were grasping a normal object. We are currently working to undertake more thorough HCI experiments to evaluate the Figure 10 gloves compared to the standard wireless controllers shown in Figure 1. In separate set of experiments, we have devised a sonification algorithm which lets participants hear a real-time feed of the nonbonded protein-ligand energies in section 4.4, and evaluating the extent to which this audio feed helps users to find drug binding poses, and carry out binding and rebinding tasks.

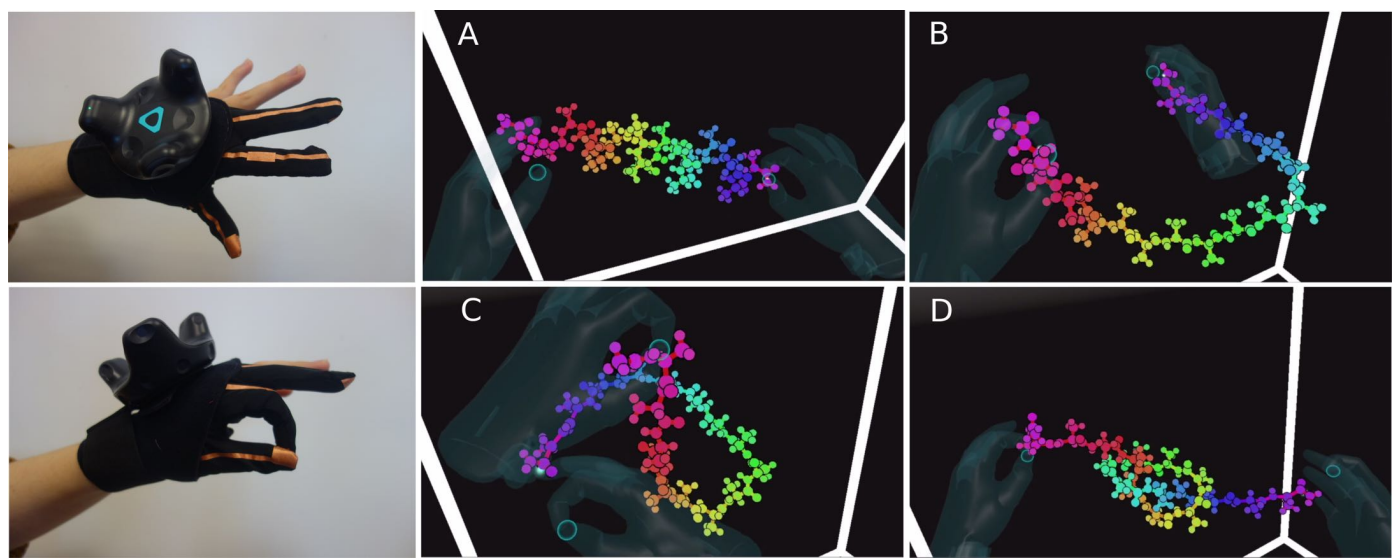

Figure 10: images on the left show the pinch sensing data glove we designed to enable robust molecular manipulation. The sequence of images on the right (A – D) show a glove-wearing iMD-VR user's first-person perspective within Narupa as they tie a knot in a real-time simulation of a 17-ALA peptide. A and B show the user unravelling the helical secondary structure; C and D show the protein being tied into a knot.

## 6. Conclusions & Future Directions

In this article, we have attempted to provide an overview of possibilities which the current generation of immersive technologies (spanning both VR and XR) hold for advancing the molecular sciences. We have introduced our open-source multi-person iMD-VR framework Narupa, and described some of our initial applications across different areas of molecular science, including small molecules, materials, and biochemistry. The applications which we have outlined in this paper, obtained by connecting our iMD-VR client to both molecular mechanics and quantum mechanics force engines – are encouraging. For studying 3d conformational dynamics, the results in section 4.1 shown that iMD-VR enables efficiency gains of 2 – 10x compared to standard computer interfaces. The results in section 4.2 demonstrate that iMD-VR offers an efficient way to train machines to learn reactive potential energy surfaces. Curating accurate training data remains a longstanding problem for learning PES topologies; the paper by Amabilino *et al.* shows how data curated (in one hour by a single graduate student!) using iMD-VR to steer 'on-the-fly' *ab* initio dynamics can be used to train neural nets which are accurate in the vicinity of the minimum energy path. The results in section 4.3 show that studies like those carried out Amabilino *et al.* can be utilized as kind of more generalized search strategy to discover important molecular conformations and transition pathways on reactive PESs, similar to the observation that arose from the FOLDIT studies: humans are extremely adept at solving complex spatial search problems.[97] Section 4.4 (exploring loop motion in CypA) and section 4.5 (outlining iMD-VR drug docking studies) provide two important insights into the results which participants generate using iMD-VR, illustrating that careful participants can generate results which are both reversible and reproducible. Combined with the results of Amabilino *et al.*, *these observations suggest that the regime which participants sample during the course of an iMD-VR run does not stray too far from the equilibrium ensemble*. Section 4.6 highlights an exciting emerging application of iMD-VR, which is important in a wide range of industrial contexts, because it will enable us to get a sense of the extent to which pores can be engineered in zeolite structures so as to tune molecular transport properties. Together, these applications provide strong evidence that our iMD-VR framework provides the sophistication required for experts to make progress on complex and bespoke scientific projects – i.e., it satisfies the 'high-ceiling' design criterion which we discussed in the introduction.

Our observations to date suggest that new tools like iMD-VR have the potential to change the kind of science that people undertake,[115] but there remains a great deal of work to be done in developing the technology and understanding those applications to which it is best suited. The ideal scenario is one in which technology development and scientific applications are closely coupled, so that each can inform the other. In the near term, we believe that the most promising applications for tools like iMD-VR are those which: (1) are intrinsically three-dimensional and (2) require exploring extremely high dimensional search spaces. For these sorts of problems, brute force strategies take a very long time to converge, and moreover the human brain is quite adept at using its powers of spatial reasoning to make intelligent mechanistic and design hypotheses, especially when presented with interfaces that enable its spatial reasoning powers to be fully expressed – i.e., without wasting valuable time fiddling with a 2d molecular visualization tool, or a command line scripting interface.

Exploring 3D conformational dynamics on rugged PESs to determine kinetic transition rates between minima on a PES, and also providing free energies and kinetics along those pathways is an "NP-complete" problem for which no optimal method exists,[116] and a good example of the kind of problem where we plan to apply iMD-VR. For PES searching, brute force methods are impossible for all but the smallest systems, and sophisticated search algorithms often struggle (e.g., umbrella sampling, adaptive force biasing, temperature accelerated MD, metadynamics,[117] replica exchange, transition path sampling, string methods, markov state models,[104, 118] forward flux sampling, milestoning, adaptive BXD,[119] etc.). Most of these methods require as input an initial guess at a pathway, or else a set of collective variables (CVs) along which biasing should be carried out, and the quality of the results they generate depends on how close the initial pathway or CV definition is to the actual minimum free energy path (MFEP) a system takes as it moves between minima. iMD-VR allows molecular scientists to combine spatial reasoning, proprioceptive 'feel' of the molecular system, auditory renderings of observables like the potential energy, and their molecular training to effectively 'draw' candidate pathways within the 3N-dimensional simulation space. The results in section 4.1 (threading $CH_4$ through a nanotube, changing helicene chirality, and tying a peptide knot), section 4.4 (loop motion in CypA) and section 4.5 (drug docking) represent our first attempts toward the goal of allowing molecular scientists to sketch dynamical hypotheses in 3D, and provide compelling evidence that iMD-VR participants are indeed able to quickly generate good pathways. In order to derive a free energy from these 3D drawings, an additional post-processing step is required – namely, the 3N-dimensional path must be processed using a dimensionality reduction algorithm, which can then be used as input to a path-based free energy sampling method. Whilst we have made progress toward developing additional tools which enable iMD-VR to form a workflow for path-based free energy sampling, further methods development efforts are required to enable this idea to become a reality which can be applied to a wide range of systems.

We are currently making progress in developing a VR-enabled molecular builder which can be used in conjunction with our iMD-VR framework. This will give molecular designers the ability to atomistically modify molecular structures during iMD-VR simulation runs, and evaluate the dynamical and functional consequences of structural modifications. For this unified builder-simulator framework, we envision molecular machines and synthetic biology to be two important application domains. The 2016 Nobel prize in chemistry to Sauvage, Stoddart, and Feringa highlighted molecular machines with controllable movements which can perform tasks when energy is added. Molecular machines span a wide range of application domains – e.g., molecular walkers,[120] molecular pumps,[121] molecular information ratchets,[122] molecules that can synthesize other molecules,[123] interlocked molecular rotors,[124] nanocars,[125] and many others. From an iMD-VR perspective molecular machines represent an interesting application area because: (1) the potential design space for molecular machines is enormous; (1) their function is intrinsically dynamic, requiring the directed injection of energy into the system; and (3) the majority of the machines reported to date have structures and dynamics which are intrinsically three dimensional, and therefore challenging for conventional 2D modelling interfaces. These same principles are also true within synthetic biology,[126] where researchers seek to use molecular design in order to develop targeted drug therapies, augment photosynthesis,[127] construct protein architectures which can operate as purpose-built catalysts,[128, 129] or develop new supramolecular structures which can bind small molecules.[130] For this reason, we are looking to extend Narupa's flexible API beyond atomistic modelling to also communicate with coarse-grained modelling engines.

Like many domains of scientific computing, the basic workflow for molecular simulation has remained largely unchanged for the last 30 – 40 years: i.e., iterative cycles of job submission to HPC resources, followed by visualization on a 2D display.[8] At some point, this paradigm will change, and it may be that immersive technologies like VR, combined with the power of modern HPC and fast networks, drive this change. The extent to which a new technology ends up being adopted within a particular domain is difficult to foresee; nevertheless, we believe that the range of research applications outlined herein provide a glimpse into what might be possible should next-generation immersive interaction technologies like iMD-VR find more widespread use within the molecular sciences. Adoption is only likely to arise by demonstrations (e.g., controlled user studies) which show that XR technologies are better than existing technologies in some measurement space, and also by good applications which generalize to other areas of nanoscience.

With the ever-accelerating rise of machine learning and automation across all areas of science and society, immersive technologies like iMD-VR represent an interesting research domain precisely because they cannot be disentangled from issues linked to human perception. The usage and development of iMD-VR tools represents an inherently cross-disciplinary pursuit, which – if it is to be successful – must connect scientists, technologists, interaction designers, artists, and psychologists. With recent advances in high-performance computing, data science, robotics, and machine learning, many have begun to speculate about the future of scientific practice, asking important questions as to the sort of scientific future we should be consciously working to design over the next few decades.[23, 51] In an increasingly automated future which is reliant on machines, it is important to think carefully about and discuss the role which human creative expression and human perception will play. Narratives of our emerging technological future sometimes default to a philosophical sentiment which casts automation as the ultimate end, sometimes leaving one to wonder where exactly the human fits in. So long as human creativity continues to play an important role in the process of scientific understanding, discovery, and design, then we believe that immersive frameworks like iMD-VR may have a crucial role to play in our emerging scientific future. *Precisely because iMD-VR is a technology which is ultimately designed for the human perceptual system, it represents a technology where the human cannot be automated away*. In our view, advanced visualization and interaction frameworks are *complementary* to research activities aimed at increasing the automation of research tasks and scientific discovery, because they provide an efficient means for humans to undertake communication and collaboration, and express high-level creative scientific and design insight, leaving automated frameworks to subsequently sort out the computational and mechanistic details. The recent paper by Amabilino et al.[13] offers an example of how human experts can utilize iMD-VR to train ML in order to accelerate molecular science workflows. In the near term, ML will not eliminate humans; rather it will result in a scenario where humans focus their efforts on a different kind of problem: how to best train a machine. Amabilino et al. gives one particular example of what might be possible moving forward, and how VR might be productively used in conjuction with ML to accelerate scientific workflows.

At the moment, computational science tends to privilege those who are able to deftly process a particular flavor of mathematical cognitive abstraction. One interesting comment we have encountered to our iMD-VR work is that it makes things 'too easy', and risks researchers feeling they no longer need to toil away in the details of mathematical abstraction. However, in some sense this is the definition of scientific progress. For example, there was a time when anybody who wanted to solve a matrix eigenvalue problem wrote their own diagonalizer. Nowadays most scientists are content to use an existing diagonalization tools, so they can focus their intellectual energy on other kinds of problems. Technology like iMD-VR enables us to transform intangible mathematical abstractions like molecular force fields into more tangible realities, which consequently engage a broader range of our sensory modalities. In so doing, we can make complicated problems more accessible to a wider spectrum of intelligences, perhaps facilitating creative solutions which have been heretofore inaccessible.

**Acknowledgements**

MOC has received support through the Royal Society (RGF\EA\181075), BBSRC (BB/R00661X/1), and an EPSRC industrial CASE studentship. SJB thanks EPSRC grant EP/M022129/1, HECBioSim, the University of Bristol School of Chemistry, and a BP Innovation Fellowship. HMD and AJB are funded by PhD studentships from EPSRC. Funding for AJB is from the EPSRC Centre for Doctoral training, Theory and Modelling in Chemical Sciences, under grant EP/L015722/1. RJS is supported by EPSRC Programme grant EP/P021123/1. TJM thanks the Leverhulme Trust, Royal Society and the Academies for funding (APX\R1\180118). AJM thanks EPSRC for funding (EP/M022609/1) and also the Collaborative Computational Project for Biomolecular Simulation (CCP-BioSim, www.ccpbiosim.ac.uk), supported by EPSRC. DRG acknowledges funding from: Oracle Corporation (University Partnership Cloud award); the Royal Society (*URF\R\180033*); EPSRC (impact acceleration award, institutional sponsorship award, and EP/P021123/1), the Leverhulme Trust (Philip Leverhulme Prize); and the London Barbican (Open Lab Funding). We also thank the following: Sidd Khajuria (Barbican), Chris Sharp (Barbican), James Upton (Royal Society), Simon McIntosh-Smith (University of Bristol), Dek Woolfson (University of Bristol), Lisa May Thomas (University of Bristol), Rachel Freire (London), Becca Rose Glowacki (University of the West of England); and Interactive Scientific (Bristol) for support and engagement throughout this work; Jenny Tsai-Smith (Oracle) and Phil Bates (Oracle) for provision of cloud computing facilities; Markus Reiher, Alain C. Vaucher, and Thomas Weymouth (ETH) for support designing a plugin between Narupa and the SCINE code; Julien Michel and Jordi Juarez Jimenez (Edinburgh) for helpful discussions about the Cyclophilin-A system; Anders S. Christensen, Jimmy C. Kromann, and O. Anatole von Lilienfeld (Basel) for engaging with us to build a plugin between Narupa and the SCINE code; and Ed Clarke and Ellie Burfoot for their help in carrying out studies evaluating reaction discovery using quantum mechanical force engines.

# References

[1] Krueger, in *Proceedings of the June 13-16, 1977, national computer conference* (ACM, Dallas, Texas, 1977), pp. 423;
[2] Talsma, Frontiers in Integrative Neuroscience 9 (2015);
[3] Proulx et al., Neuroscience & Biobehavioral Reviews 41, 16 (2014);
[4] Glowacki et al., Faraday Discuss. 169, 63 (2014);
[5] Mitchell et al., Leonardo 49, 138 (2016);
[6] Arbon et al., in *Proceedings of the Interntional Conference on Audio Displays* (2018), Vol. 30;
[7] Davies et al., in *Evolutionary and Biologically Inspired Music, Sound, Art and Design. EvoMUSART 2016*, edited by C. Johnson, V. Ciesielski, J. Correia and P. Machado (Springer International Publishing, Cham, 2016), Vol. 9596, pp. 17;
[8] O'Connor et al., Science Advances 4, eaat2731 (2018);
[9] Thomas, Glowacki, International Journal of Performance Arts and Digital Media 14, 145 (2018);
[10] Cruz-Neira, Sandin, DeFanti, in *Proceedings of the 20th annual conference on Computer graphics and interactive techniques* (ACM, Anaheim, CA, 1993), pp. 135;
[11] Seymour et al., Annals of surgery 236, 458 (2002);
[12] Wallen, in *Supercomputing 2014* (New Orleans, 2014);
[13] Amabilino et al., The Journal of Physical Chemistry A, DOI: 10.1021/acs.jpca.9b01006 (2019);
[14] Weibel, Fruk, *Molecular Aesthetics*. (MIT Press, 2013);
[15] Sutherland, Proceedings of the IFIP Congress 2, 506 (1965);
[16] Feynman, in *Feynman and computation*, edited by J. G. H. Anthony (Perseus Books, 1999), pp. 63;
[17] Fuechsle et al., Nature Nanotechnology 7, 242 (2012);
[18] Leinen, Beilstein Journal of Nanotechnology 6, 2148 (2015);
[19] Gibson, *The senses considered as perceptual systems*. (Houghton Mifflin, Boston, 1966);
[20] Gibson, *The Ecological Approach to Visual Perception*. (Houghton Mifflin, Boston, 1979);
[21] Withagen et al., New Ideas in Psychology 30, 250 (2012);
[22] Norman, interactions 6, 38 (1999);
[23] Aspuru-Guzik, Lindh, Reiher, ACS Central Science 4, 144 (2018);
[24] Goddard et al., Journal of Molecular Biology 430, 3982 (2018);
[25] Ai, Fröhlich, Computer Graphics Forum 17, 267 (1998);
[26] Anderson, Weng, Journal of Molecular Graphics and Modelling 17, 180 (1999);
[27] Moritz, Meyer, in *Proceedings of the Fourth IEEE Symposium on Bioinformatics and Bioengineering* (2004), pp. 503;
[28] Férey et al., Virtual Reality 13, 273 (2009);
[29] Block et al., Source Code for Biology and Medicine 4, 3 (2009);
[30] Au - Doblack, Au - Allis, Au - Dávila, JoVE, e51384 (2014);
[31] Balo, Wang, Ernst, Nature Methods 14, 1122 (2017);
[32] García-Hernández, Kranzlmüller, in *Augmented Reality, Virtual Reality, and Computer Graphics. AVR 2017. Lecture Notes in Computer Science* (Springer International Publishing, Cham, 2017), Vol. 10324, pp. 309;
[33] Borrel, Fourches, Bioinformatics 33, 3816 (2017);
[34] Stone, Sherman, Schulten, in *2016 IEEE International Parallel and Distributed Processing Symposium Workshops (IPDPSW)* (2016), pp. 1048;
[35] Zheng, Waller, Journal of Molecular Graphics and Modelling 73, 18 (2017);
[36] Norrby et al., Journal of Chemical Information and Modeling 55, 2475 (2015);
[37] Grebner et al., Future Medicinal Chemistry 8, 1739 (2016);
[38] Salvadori et al., International Journal of Quantum Chemistry 116, 1731 (2016);
[39] Ratamero et al., Journal of Computer-Aided Molecular Design 32, 703 (2018);
[40] Myers et al., ACM Trans. Comput.-Hum. Interact. 7, 3 (2000);
[41] Wessel, Wright, Comput. Music J. 26, 11 (2002);
[42] Hyde, Mitchell, Glowacki, in *Proceedings of the 17th International Generative Art Conference, (GENArt 2014)* (Roma, Italia, 2014);
[43] Slater, Sanchez-Vives, Frontiers in Robotics and AI 3 (2016);
[44] Allinger, Yuh, Lii, Journal of the American Chemical Society 111, 8551 (1989);
[45] Eastman et al., Journal of Chemical Theory and Computation 9, 461 (2013);
[46] Aradi, Hourahine, Frauenheim, The Journal of Physical Chemistry A 111, 5678 (2007);

**[47]** Husch, Vaucher, Reiher, International Journal of Quantum Chemistry 118, e25799 (2018);
**[48]** Bonomi et al., Computer Physics Communications 180, 1961 (2009);
**[49]** Todorov et al., Journal of Materials Chemistry 16, 1911 (2006);
**[50]** Becca Rose Glowacki, arXiv:1901.03532 [cs.HC] (2019);
**[51]** Lanier, *Dawn of the new everything: a journey through virtual reality.* (Penguin, London, 2017);
**[52]** Swapp, Pawar, Loscos, in *Virtual Reality* (2006), Vol. 10, pp. 24;
**[53]** Crowfoot et al., in *Chemistry of penicillin*, edited by H. T. Clarke, J. R. Johnson and R. Robinson (Princeton University Press, Princeton, New Jersey, 1949), pp. 310;
**[54]** Pauling, Corey, Branson, Proceedings of the National Academy of Sciences 37, 205 (1951);
**[55]** Crick, Watson, Proc. Royal Soc. A 223, 80 (1954);
**[56]** Kendrew et al., Nature 181, 662 (1958);
**[57]** Perutz et al., Nature 185, 416 (1960);
**[58]** Brooks et al., ACM SIGGraph computer graphics 24, 177 (1990);
**[59]** Atkinson et al., Computers & Graphics 2, 97 (1977);
**[60]** Surles et al., Protein Sci 3, 198 (1994);
**[61]** Brooks, Faraday Discuss. 169, 521 (2014);
**[62]** Ming, Beard, Brooks, in *IEEE Int. Conf. on Robotics and Automation* (IEEE, 1989), pp. 1462;
**[63]** Bejczy, Science 208, 1327 (1980);
**[64]** Pollack, VMD: 20 Yrs History and Innovation, available at ks.uiuc.edu/History/VMD/;
**[65]** Dreher et al., Procedia Comp. Sci. 18, 20 (2013);
**[66]** Haag et al., ChemPhysChem 15, 3301 (2014);
**[67]** Luehr, Jin, Martínez, Journal of chemical theory and computation 11, 4536 (2015);
**[68]** Slater, Frontiers in Robotics and AI 1 (2014);
**[69]** Thomas et al., arXiv:1901.03536 [cs.HC] (2019);
**[70]** Cho et al., Computer Methods and Programs in Biomedicine 113, 258 (2014);
**[71]** Stone, Gullingsrud, Schulten, in *Proceedings of the 2001 symposium on Interactive 3D graphics* (ACM, 2001), pp. 191;
**[72]** Glowacki et al., (fOOm-d, a framework for object-oriented molecular dynamics, https://sourceforge.net/projects/foom-d/);
**[73]** Plimpton, Journal of Computational Physics 117, 1 (1995);
**[74]** Berendsen et al., The Journal of Chemical Physics 81, 3684 (1984);
**[75]** Miyamoto, Kollman, J Comput Chem 13, 952 (1992);
**[76]** Eastman, Pande, Journal of Chemical Theory and Computation 6, 434 (2010);
**[77]** Kabsch, Sander, Biopolymers 22, 2577 (1983);
**[78]** Kalra, Hummer, Garde, The Journal of Physical Chemistry B 108, 544 (2004);
**[79]** De Poli et al., Science 352, 575 (2016);
**[80]** Alder, Butts, Sessions, Chem Sci 8, 6389 (2017);
**[81]** Gil-Ramírez et al., Journal of the American Chemical Society 138, 13159 (2016);
**[82]** Ziegler et al., Proc. Natl. Acad. Sci. U. S. A. 113, 7533 (2016);
**[83]** Dunning et al., Science 347, 530 (2015);
**[84]** Glowacki, Orr-Ewing, Harvey, The Journal of Chemical Physics 143, 044120 (2015);
**[85]** Haag, Reiher, International Journal of Quantum Chemistry 113, 8 (2012);
**[86]** Vaucher, Haag, Reiher, Journal of Computational Chemistry 37, 805 (2016);
**[87]** Haag, Reiher, Faraday Discussions 169, 89 (2014);
**[88]** Christensen et al., (http://www.qmlcode.org);
**[89]** Gao et al., Computer Physics Communications 203, 212 (2016);
**[90]** Varela, Vazquez, Martinez-Nunez, Chem Sci 8, 3843 (2017);
**[91]** Martinez-Nunez, J Comput Chem 36, 222 (2015);
**[92]** Wang et al., Nat Chem 6, 1044 (2014);
**[93]** Zador, Miller, P Combust Inst 35, 181 (2015);
**[94]** Zheng, Pfaendtner, J Phys Chem C 118, 10764 (2014);
**[95]** Shannon et al., Journal of Chemical Theory and Computation 14, 4541 (2018);
**[96]** Ohno, Maeda, Chemical Physics Letters 384, 277 (2004);
**[97]** Cooper et al., Nature 466, 756 (2010);
**[98]** Eiben et al., Nat Biotechnol 30, 190 (2012);

**[99]** Heck et al., Proc Natl Acad Sci U S A 115, E11231 (2018);
**[100]** Eisenmesser et al., Nature 438, 117 (2005);
**[101]** Chi et al., Angewandte Chemie International Edition 54, 11657 (2015);
**[102]** Plattner, Noé, Nature Communications 6, 7653 (2015);
**[103]** Buch, Giorgino, De Fabritiis, Proceedings of the National Academy of Sciences 108, 10184 (2011);
**[104]** Wu et al., Proceedings of the National Academy of Sciences 113, E3221 (2016);
**[105]** Huggins et al., Wiley Interdisciplinary Reviews: Computational Molecular Science 0, e1393 (2018);
**[106]** Yilmaz, Müller, Topics in Catalysis 52, 888 (2009);
**[107]** Czaja, Trukhan, Müller, Chemical Society Reviews 38, 1284 (2009);
**[108]** Bu et al., Catalysis Today 312, 73 (2018);
**[109]** Granato et al., Chemical Engineering Science 65, 2656 (2010);
**[110]** Bereciartua et al., Science 358, 1068 (2017);
**[111]** Rau et al., in *2015 IEEE 1st International Workshop on Virtual and Augmented Reality for Molecular Science (VARMS@IEEEVR)* (2015), pp. 25;
**[112]** Grand, Dall Antonia, in *Proceedings of the Interntional Conference on Audio Displays* (2008), Vol. 30;
**[113]** Ballweg, Bronowska, Vickers, in *Proceedings of ISon 2016, 5th Interactive Sonification Workshop, CITEC* (2016);
**[114]** Shams, Kamitani, Shimojo, Nature 408, 788 (2000);
**[115]** Brooks, Faraday Discuss. 169, 521 (2014);
**[116]** Hart, Istrail, J. Comput. Biol. 4, 1 (1997);
**[117]** Pérez de Alba Ortíz et al., The Journal of Chemical Physics 149, 072320 (2018);
**[118]** Husic, Pande, Journal of the American Chemical Society 140, 2386 (2018);
**[119]** O'Connor et al., Faraday Discuss. 195, 395 (2016);
**[120]** von Delius, Geertsema, Leigh, Nature Chemistry 2, 96 (2009);
**[121]** Cheng et al., Nature Nanotechnology 10, 547 (2015);
**[122]** Serreli et al., Nature 445, 523 (2007);
**[123]** Kassem et al., Nature 549, 374 (2017);
**[124]** Leigh et al., Nature 424, 174 (2003);
**[125]** Kudernac et al., Nature 479, 208 (2011);
**[126]** Grayson Katie, Anderson, Journal of The Royal Society Interface 15, 20180472 (2018);
**[127]** Swift Thomas et al., Interface Focus 9, 20180048 (2019);
**[128]** Röthlisberger et al., Nature 453, 190 (2008);
**[129]** Watkins et al., Nature Communications 8, 358 (2017);
**[130]** Thomas et al., ACS Synthetic Biology 7, 1808 (2018);